\documentclass[times]{TRR}

\usepackage{moreverb,url}
\usepackage{multirow}
\usepackage{physics, amsmath, float}
\usepackage{verbatim} 

\newcommand\BibTeX{{\rmfamily B\kern-.05em \textsc{i\kern-.025em b}\kern-.08em
T\kern-.1667em\lower.7ex\hbox{E}\kern-.125emX}}

\begin{document}

\title{Sustainability of Stack Exchange Q\&A communities: the role of trust}

\author{Ana Vrani\'c\affilnum{1}, Aleksandar Toma\v{s}evi\'c \affilnum{2}, Aleksandra Alori\'c\affilnum{1,3} and Marija Mitrovi\'c Dankulov\affilnum{1}}

\affiliation{\affilnum{1}Institute of Physics Belgrade, University of Belgrade, Pregrevica 118, 11080 Belgrade, Serbia\\
\affilnum{2}Department of Sociology, Faculty of Philosophy, University of Novi Sad, Serbia\\
\affilnum{3}Two desperados, Serbia}


\begin{abstract}
Knowledge-sharing communities are a fundamental element of any knowledge-based society. Understanding how they emerge, function, and disappear is thus of crucial importance. Many social and economic factors influence sustainable knowledge-sharing communities. Here we explore the role of the structure of social interactions and social trust in the emergence of these communities. Using tools from complex network theory, we analyze the early evolution of social structure in four pairs of StackExchange communities, each corresponding to one active and one closed community on the same topic. We adapt the dynamical reputation model to quantify the evolution of social trust in these communities. Our analysis shows that active communities have higher local cohesiveness and develop stable and more strongly connected cores. The average reputation is higher in sustainable communities. In these communities, the trust between core members develops early and remains high over time. Our results imply that efforts to create a stable and trustworthy core may be crucial for building a sustainable knowledge-sharing community. 
\end{abstract}

\maketitle

\section{Introduction}

The development of a knowledge-based society is one of the critical processes in the modern world \cite{leydesdorff2001sociological,leydesdorff2012triple}. In a knowledge-based society, knowledge is generated, shared, and made available to all members. It is a vital resource. Sharing this resource between individuals and organizations is a necessary process, and knowledge-sharing communities are one of the fundamental elements of a knowledge society.\\

Often, these knowledge-sharing communities depend on the willingness of their members to engage in an exchange of information and knowledge. Participation in the community is voluntary, with no noticeable material gains for members. Thus, the exchange of knowledge depends on mutual trust between members. Trust that the community will consider their questions essential for the growth of the knowledge corpus and invest resources to answer their questions. Trust that the community will objectively evaluate their answers based on their quality and clarity. The trust mentioned is beyond the direct trust formed between two members. It is a feeling of a member that a community can be trusted and that their engagement is valuable. A feeling of community that a member can be trusted and expressed through engaging that member in community activities. It is a collective phenomenon that depends on and is built through social interactions between community members. This is why we believe it is crucial to understand how trustworthy knowledge-sharing communities emerge and disappear, as well as to unveil the fundamental mechanisms that underlie their evolution and determine their sustainability.\\

In the past two decades, we have witnessed the emergence of an online knowledge-sharing community StackOverflow, which has become one of the most popular sites in the world and the primary knowledge resource for coding. The success of StackOverflow led to the emergence of similar communities on various topics and formed the StackExchange (SE) network.\footnote{
 More information about StackOverflow is available at: \url{https://stackoverflow.co/} and broad introduction to StackExchange (SE) network is available at: \url{https://stackexchange.com/tour}. Visit \url{https://area51.stackexchange.com/faq} for more details about closed and beta SE communities and the review process.
 } 
The advancement of Information and communication technologies (ICTs) have enabled faster and easier creation and sharing of knowledge, but also the access to a large amount of data that allowed a detailed study of their emergence and evolution \cite{dankulov2015dynamics}, as well as user roles \cite{saxena2021users}, and patterns of their activity \cite{santos2019activity, slag2015one, chhabra2020activity}. However, relatively little attention has been paid to the sustainability of SE communities. Most research focused on the activity and factors that influence the users' activity in these communities. Factors such as the need for experts and the quality of their contributions have been thoroughly investigated \cite{dev2018size}. It was shown that the growth of communities and mechanisms that drive it might depend on the topic around which the community was created \cite{santos2019self}. \\

\begin{figure*}
    \centering
    \includegraphics[width=0.9\linewidth]{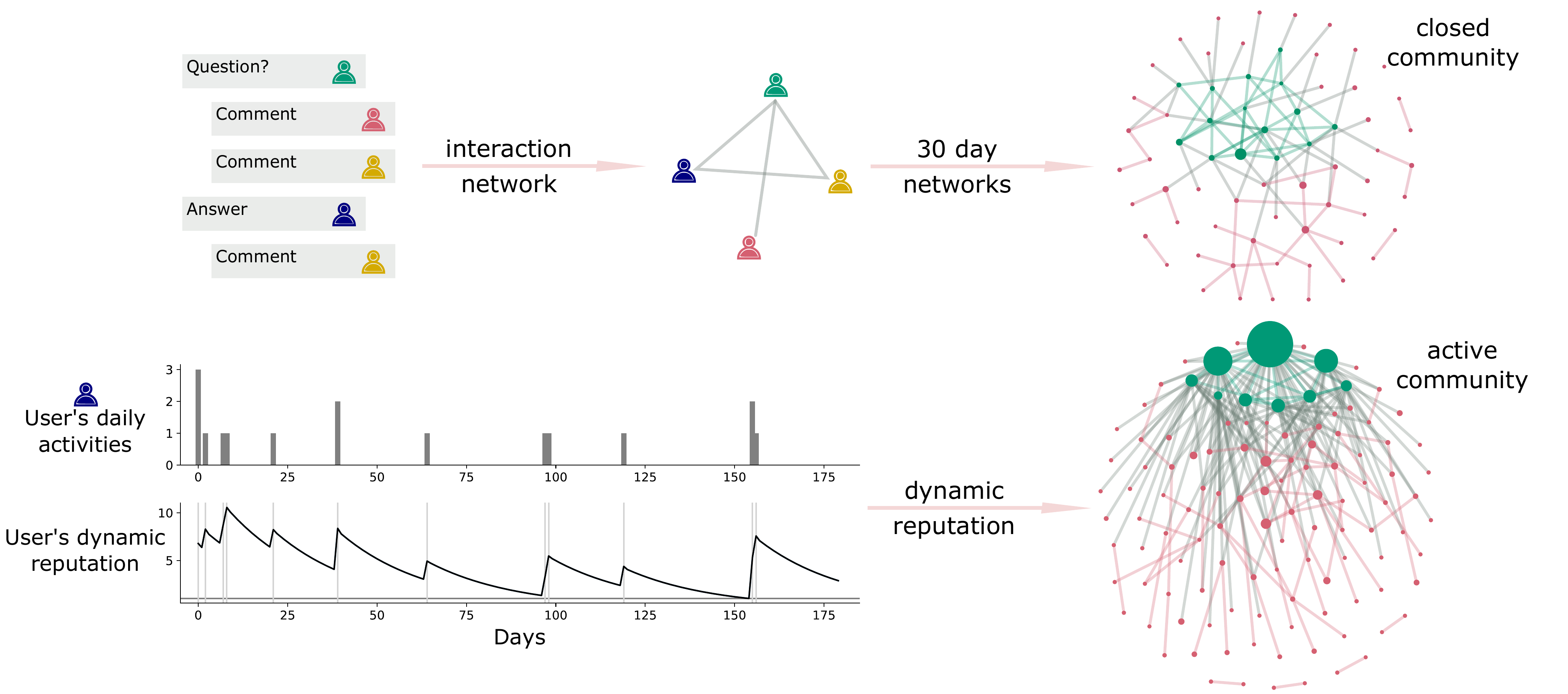}
    \caption{\textbf{Visual abstract:} Top row illustrates how user interaction via questions, answers, and comments is translated into an undirected network of interactions between users and finally aggregated over 30 day windows. The bottom row shows activity and corresponding dynamic reputation for one user from the closed literature SE community. Networks on the right illustrate differences between closed and active SE literature communities. Nodes are colored according to the core/periphery affiliation, while their size corresponds to dynamic reputation on the last day of interaction that the network contains.}
    \label{fig:paper_summary}
\end{figure*}

In this paper, we investigate the role of network structure and social trust in the sustainability of a knowledge-sharing community. Research on the sustainability of social groups shows that social interaction and their structure influence the dynamics and sustainability of social groups \cite{oliver2001whatever, smiljanic2017associative, torok2017cascading, lHorincz2019collapse}. Dyadic trust between members may not play an essential role in the group dynamics of knowledge-sharing communities. However, it is known that the reputation of users, one of the proxies of trust in online communities, is the primary motivating factor for them to become and maintain their productive member status \cite{waskoWhyShouldShare2005, hungInfluenceIntrinsicExtrinsic2011,rodeShareNotShare2016}. The reputation helps users manage the complexity of the collaborative environment by signaling out trustworthy members. \\

With the proliferation of misinformed decisions, it is crucial to understand how to foster communities that promote collaborative knowledge exchange and understand how cooperative norms of trustworthy behavior emerge. The way people interact, specifically the structure of their interactions \cite{kairam2012life}, and how inclusive and trustworthy the key members of the community may influence the sustainability of the knowledge-sharing communities. Although the topic and early adopters are essential in establishing a new SE community, they are not sufficient for sustainability. The current SE network has several examples of communities whose establishment was unsuccessful at first, while the subsequent attempt resulted in a sustainable community. These pairs of unsuccessful/successful attempts allow us to investigate the relevance of social network structure and social trust in the sustainability of knowledge-sharing communities. They are particularly relevant if we wish to understand why some communities established themselves in their second attempt. For those pairs of communities, the topic is the same, and all the initial SE platform requirements were satisfied, but something else was crucial for community decay in the first attempt and its success in the second.\\

Our methods and key results are summarised in a visual abstract in Fig.~\ref{fig:paper_summary}. We analyze four pairs of SE communities and study the differences in the evolution of social structure and trust between closed and active communities. We have selected four topics from the STEM and humanities: astronomy, physics, economics, and literature. We focused on the topics where we could find a matched pair of closed and active communities. For this reason alone, we do not include StackOverflow as the most popular community in our analysis. We analyze each pair's early stages of evolution and look at the differences between active and closed communities. Specifically, we map the interactions onto complex networks and examine how their properties evolve during the first 180 days of communities' existence. Using complex network theory \cite{boccaletti2006complex} we quantify structure of these networks, and compare their evolution in active and closed communities on the same topic. We pay special attention to the core-periphery structure of these networks since it is one of the most prominent features of social networks \cite{gallagher2020clarified}. We examine how core-periphery structure of active and closed communities evolve and analyze their difference. We show that active communities have a higher value of local clustering and a more stable core membership. On average, the core of the sustainable communities has higher inner connectivity and is more connected with the periphery.\\

To study the evolution of social trust, we adapt the Dynamic Interaction Based Reputation Model (DIBRM) \cite{melnikovDynamicInteractionBasedReputation2018}. The model allows us to quantify the trust of each individual over time. We can quantify members' mean and total trust within the core and periphery and follow their evolution through time. The mean reputation of members is higher in sustainable communities than in closed ones, indicating higher levels of social trust. Furthermore, the mean reputation of core members of active communities is constantly above the mean reputation of core members in closed communities, indicating that the creation of trust in the early stages of a community's life may be crucial for its survival. Our results show that social organization and social trust in the early phases of the life of a knowledge-sharing community have an essential role in its sustainability. Our analysis reveals differences in the evolution of these properties in communities on different topics. 

The paper is organized as follows. In Sec. \textit{Previous research} we give a short overview of previous research. Sec. \textit{Data} describes the data and outlines some specific properties of each community. In Sec. \textit{Method} we describe the measures and models used for describing the local organization and measuring reputation. Section \textit{Results} shows our results. Finally, we discuss our results and selection of model parameters and time window, as well as its consequences in Sec. \textit{Discussion and conclusions}.

\section{Previous research}

The availability of data from the SE network led to detailed research on the different aspects of dynamics of knowledge sharing communities  \cite{dankulov2015dynamics, santos2019activity, slag2015one, chhabra2020activity}, the roles of users \cite{saxena2021users}, and their motivations to join and remain members of these communities \cite{wei2015motivating, yanovsky2019one, santos2020can, bornfeld2019interaction, kang2021motivational}. The focus of the research in the previous decade was on the evolution of activity in SE communities and the different factors that influence this growth. Ahmed et al. \cite{ahmed2015does} have investigated differences between technical and non-technical communities and showed that within the first four years, technical communities have a higher growth rate, more activity, and are more modular. The comparison of UX community in SE and Reddit \cite{chen2021characterizing} showed that the Reddit community grows faster, while SE becomes less diverse and active over time. Special attention was paid to the activities of individual users. In Ref. \cite{posnett2012mining} authors argue that while the overall quality of the answers, measured in the answer score, decays over time, the quality of the answers of the individual user remains constant. This observation suggests that good answerers are \textit{born} and not made within the community. Reputation is used as a proxy for the recognition of experts \cite{pal2012evolution} by other members. However, contrary to common sense, the authors show that the presence of experts can reduce the activity of other members \cite{pal2012evolution}. In \cite{santos2019self} authors explore the role of self-and cross excitation in the temporal development of user activity. Differences between growing and declining communities and communities on STEM and humanities topics were explored. Their results show that the early stages of growing communities are characterized by the high cross-excitation of a small fraction of popular users. In contrast, later stages exhibit strong long-term self-excitation in general and cross-excitation by casual users. It was also shown that cross-excitation with power users is more important in the humanities than in STEM communities, where casual users have a more critical role.\\

A relatively small number of papers focus on the sustainability of SE communities. In Ref. \cite{dev2018size}, authors examine SE sites through an economic lens. They analyze the relationship between content production based on the number of participants and activities and show that an increase in the number of questions (input) increases the number of answers (output). In their works, Oliveira et al. \cite{oliveira2018exchange} investigate activity practices and identify the tension between community spirit as proclaimed in SE guidance and individualistic values as in reputation measurement through focus groups and interviews.\\

Our assumption about the relevance of the structure of social networks in the sustainability of knowledge-sharing communities is supported by research on other social groups. Various factors influence the emergence \cite{dover2018emergence, han2017emergence}, the evolution, and the sustainability of the groups \cite{oliver2001whatever, kleineberg2015digital, oliver1985theory, kairam2012life}. The number of committed members \cite{oliver1985theory} and the minimal level of interdependence between members \cite{han2017emergence} are important factors for the emergence of the community. The levels of activity have an important role in the emergence and stability of social groups \cite{dover2018emergence, oliver1985theory}, while social factors, such as the size of the group, number of social contacts, or social capital, influence their emergence and collapse \cite{oliver2001whatever, smiljanic2017associative, torok2017cascading, lHorincz2019collapse}.\\

Another important branch of research of interest in the sustainability of online communities is the topic of trust. While ICTs make it easier for individuals to establish and maintain social contacts and exchange information and goods, they are also exposed to new risks and vulnerabilities. Social trust relationships, based on positive or negative subjective expectations of another person's future behavior, play an important but largely unexplored role in managing those risks. Recent works show that the vital element of trust is the notion of vulnerability in social relations, and as negative expectations of a trustee's behavior most often imply damage or harm to the trustor, decisions about which users to trust in an online community become paramount \cite{dunningTrustZeroAcquaintance2014,dunningWhyPeopleTrust2019,schilkeTrustSocialRelations2021}.\\

In communities such as SE, individuals have three sources of information to rely on when deciding to trust someone in a specific context: (1) knowledge of previous interactions, (2) expectations about future interactions, and (3) indirect information gained through a broader social network. Suppose the number of active users in such a community increases over a more extended period. In that case, the individuals have little or no history together, no direct interactions, and almost no memory of past interactions. In that case, the social network created by the community becomes a crucial source of information. Therefore, from a network perspective, trust can be the result of reputational concerns and can flow through indirect connections linking actors to one another \cite{schilkeTrustSocialRelations2021,mcevilyChapterNetworkTrust2021}.\\

In that case, users rely on reputation as a public measure of the reliability of other users active within the same community. Reputation is often quantified based on the history of behavior valued or promoted by a set of community norms and, as such, represents a social resource within the community \cite{aberer2001,duma2005dynamic,tschannen-moran2000}. Since reputation is public information, it is also an incentive. Agents with high reputation are motivated to act trustworthy in the future in order to preserve their status in the community \cite{mcevilyChapterNetworkTrust2021}. This idea is supported by psychological findings suggesting that trust is primarily motivated by effects produced by the act of trust itself, regardless of more rational or instrumental outcomes of trustworthy behavior \cite{dunningWhyPeopleTrust2019}. \\

In terms of modeling collective trust and reputation in online communities, knowledge about past behaviors can be implemented in a trust model in different ways. When estimating trust between agents in a social network, graph-based models focus on the topological information, position, and centrality of agents in a social network to estimate both dyadic and collective measures of social trust. On the other hand, interaction-based models, such as the dynamic reputation model implemented in this paper (DIBRM) \cite{melnikovDynamicInteractionBasedReputation2018} estimate trust or reputation based on the frequency and type of agent's interactions over time without taking into account the structure and topology of the interactions between different agents in a network.

\section{Data}

We focus on pairs of closed and active SE communities matched by topic. Astronomy, Literature, and Economics are currently active communities. All three communities thrived the second time they were proposed. The first attempt to create communities on these topics was unsuccessful, and they were closed within a year. We add to the comparison the early days of the Physics community and compare its evolution with the closed Theoretical physics community. The topics of these communities are not identical, but it is safe to assume that there is a high overlap in user demographics and interests. For these reasons, we treat this pair in the same manner as others.\\

The SE data are publicly available and released at regular time intervals. We are primarily interested in the activity and interaction data, which means that we extract the following information for posts (questions and answers) and comments: 1) for each post or comment, we extract its unique ID, the time of its creation, and unique ID of its creator - user; 2) for every question, we extract information about IDs of all answers to that question and ID of the accepted answer; 3) for each post, we collect information about IDs of its related comments. The data contains information about the official SE reputation of each user but only as a single value measuring the final reputation of the user on a day when the data archive was released. Due to this significant shortcoming, we do not include this information in our analysis. In SE, users can give positive or negative votes to questions and answers and mark questions as favorites. However, the data is again provided as a final score recorded at the release. Since this does not allow us to analyze the evolution of scores, we omit this data from our analysis.\\

\begin{table}
\footnotesize\sf\centering

\resizebox{\columnwidth}{!}{
\begin{tabular}{ccc|c|c|c|c|c} 
\cline{2-8}
                  & Site                        & \multicolumn{1}{c}{Status} & \multicolumn{1}{c}{First Date} & \multicolumn{1}{c}{$n_u$} & \multicolumn{1}{c}{$n_q$} & \multicolumn{1}{c}{$n_a$} & $n_c$  \\ 
\cline{2-8}
\multirow{2}{*}{} & \multirow{2}{*}{Physics}    & Closed                     & 09/14/11                       & 281                       & 349                       & 564                       & 2213   \\
                  &                             & Active                   & 08/24/10                       & 1176                      & 2124                      & 4802                      & 15403  \\ 
\cline{2-8}
\multirow{4}{*}{} & \multirow{2}{*}{Economics}  & Closed                     & 10/11/10                       & 275                       & 368                       & 458                       & 1253   \\
                  &                             & Active                      & 11/18/14                       & 648                       & 1024                      & 1410                      & 3553   \\ 
\cline{2-8}
                  & \multirow{2}{*}{Astronomy}  & Closed                     & 09/22/10                       & 336                       & 474                       & 953                       & 1444   \\
                  &                             & Active   & 09/24/13                       & 405                       & 644                       & 959                       & 2170   \\ 
\cline{2-8}
\multirow{2}{*}{} & \multirow{2}{*}{Literature} & Closed                     & 02/10/10                       & 284                       & 318                       & 523                       & 1097   \\
                  &                             & Active                       & 01/18/17                       & 478                       & 910                       & 907                       & 3301   \\
\cline{2-8}
\end{tabular}
}
{\raggedright \textit{Note:} Number of users $n_u$, number of questions $n_q$, number of answers $n_a$, number of comments $n_c$ \par}
\caption{Community overview for the first 180 days }
\label{tab:site-info}

\end{table}

All SE communities follow the same path from their creation until they are considered mature enough or closed. In a \textit{Definition} phase, a small number of SE users start by designing a community by proposing hypothetical questions about a certain topic. A successful \textit{Definition} phase is followed by a \textit{Commitment} phase. In this phase, interested users commit to the community to make it more active. The \textit{Beta} phase, which follows after the \textit{Commitment} phase, is the most important. It consists of two steps: a three-week private beta phase, where only committed users may ask/answer/comment questions, and a public beta phase when other members are allowed to join the community. The duration of the public beta phase is not limited. Depending on this analysis, there are three possible outcomes: 1) the community is considered successful, and it graduates; 2) the community is active but needs more work to graduate, which means that the public beta phase continues; 3) the community dies, and the site is closed. The community evaluation/review process is guided by simple metrics: the average number of questions per day, average number of answers per question, percentage of answered questions, total number of users and number of avid users, and average number of visits per day.\\

We study how the social network properties of these social communities and the social trust created among their members evolve during the first 180 days. The first 90 days are recognized as the minimal time a newly established community should spend in the beta phase. We investigate a period that is twice as long since closed communities were active between 180 and 210 days. Given that differences in the first few months of the life of the online community may help predict its survival and evolution \cite{dover2020sustainable}, we focus on the early evolution of SE sites. 

\begin{figure}
    \centering
    \includegraphics[width=\linewidth]{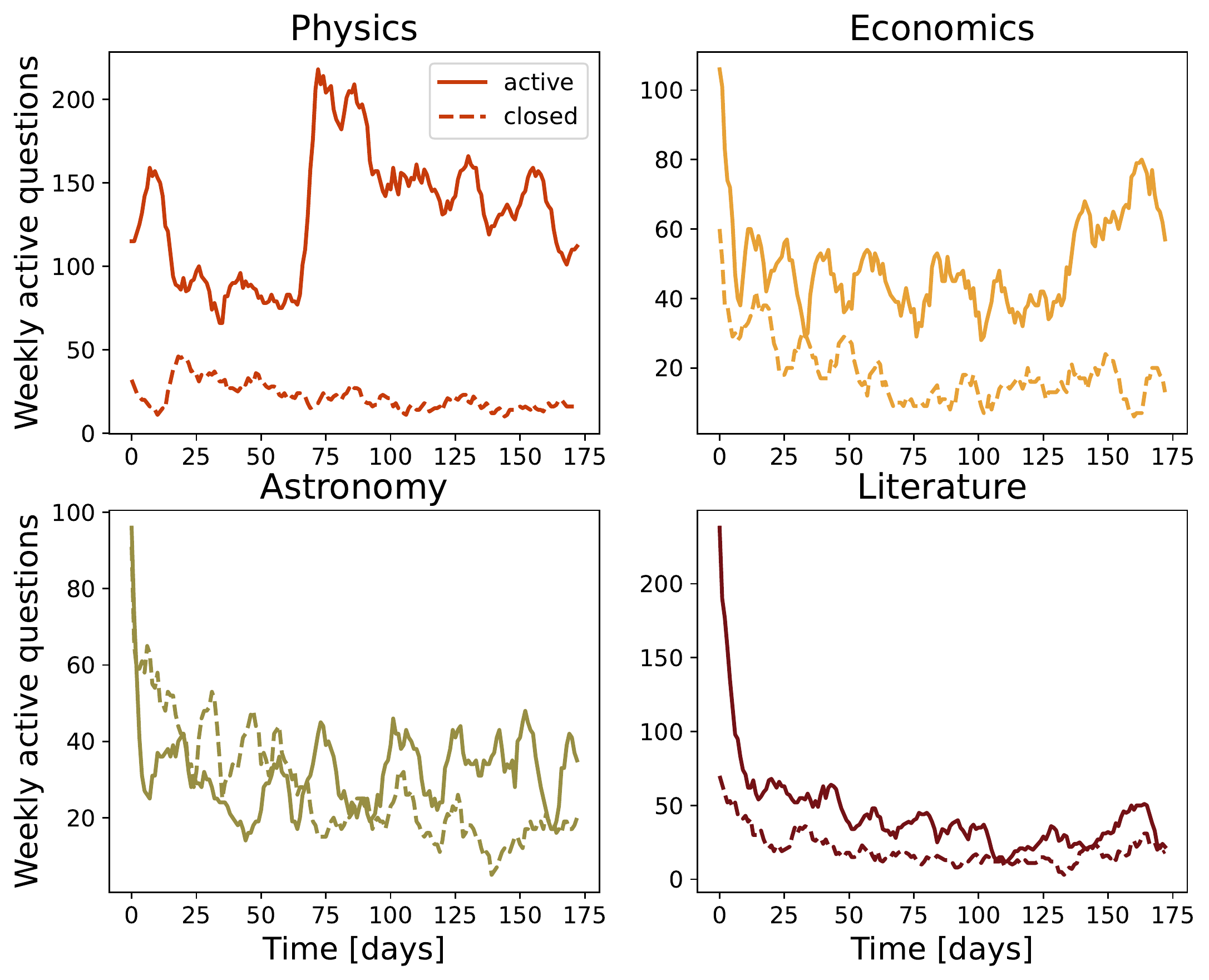}
    \caption{\textbf{Variations in the number of active questions in SE communities.} Number of active questions within 7 days sliding windows on the four pairs of Stack Exchange websites: Astronomy, Literature, Economics and Physics. Solid lines -- active sites; dashed lines -- closed sites.}
    \label{fig:active_questions}
\end{figure}

Basic information about activities collected in the first 180 days of the community life is shown in table \ref{tab:site-info}. Closed communities had fewer users, questions, and comments during this period. Although the official review of SE communities in the beta phase is based on simple activity indicators such as the number of questions or ratio of answers to questions\footnote{\url{https://stackoverflow.blog/2011/07/27/does-this-site-have-a-chance-of-succeeding/}}, these simple metrics cannot provide insight about factors which influence the success of any given community. Table A1 in Supplementary Information (SI) shows the values of some of these measures at 180 days point for considered communities. Although the Physics community was more successful than the Theoretical Physics and other considered communities, we see that these differences are not as apparent if we compare the remaining three pairs of communities. For instance, some of the parameters for the closed Astronomy community, for example, the percentage of answered questions and answer ratio, were better than for the community that is still active. Another simple indicator can be the time series of active questions for the 7 days shown in Fig. \ref{fig:active_questions}. The question is considered active if it had at least one activity, posted answer, or comment, during the previous 7 days. The four pairs of compared communities show that active communities have a higher number of active questions after 180 days. Although this difference is evident for the Physics and Economics community, Fig. \ref{fig:active_questions} shows that its value is smaller for Astronomy and Literature. Furthermore, in the case of Astronomy, the closed community had a higher number of active questions in the first 75 days. \\

The values of the measures shown in tables \ref{tab:site-info} and A1 in SI, and Fig. \ref{fig:active_questions} suggest that these simple measures are not good indicators of the long-term sustainability. Thus, we need more profound insight into the structure and dynamics of the community to understand the factors behind its sustainability. All communities must start with the same number of interesting questions, the same number of committed users, and satisfy the same thresholds to enter the public beta phase. These basic aggregated statistics are not enough to differentiate between active and closed communities. Hence, other factors determine the sustainability of communities. We investigate the role of social interaction structure and the dynamics of collective trust in the sustainability of SE communities.

\section{Method \label{sec:method}}

We are interested in the position of trustworthy members in SE communities and how active and closed communities differ regarding this factor. First, we map the interaction data onto networks and analyze their properties and how they evolve during the first 180 days. Furthermore, we use the dynamical reputation model to estimate the trustworthiness of each member of the community and the dynamics of collective trust by studying the evolution of the mean value of reputation in the community. The entire analysis was done in Python, and the whole code for reproducing results and figures is publicly available in an online repository \footnote{\url{https://anonymous.4open.science/r/Stack-Exchange-communities-C111/}}. \\

\subsection{Network mapping}

We treat all user interactions, answering questions, posting questions or comments, and accepting answers equally. We construct a network of users where the link between two nodes, users $i$ and $j$, exists if $i$ answers or comments on the question posted by $j$ and vice versa, or $i$ comments on the answer posted by $j$ and vice versa, $i$ accepts the answer posted by user $j$. We do not consider the direction or frequency of the interaction between users $i$ and $j$; thus, the obtained networks are unweighted and undirected. \\

We create a network snapshot $G(t, t+\tau$) at the time $t$ for the time window length $\tau$. Two users $(i, j)$ are connected in a network snapshot $G(t, t+\tau$) if they have had at least one interaction during the time $[t,t+\tau]$. Our first network accounts for interaction within the first 30 days $G[0,30)$, and we slide the interaction window by one day and finish with $G[149,179)$ network. This way, we create 150 interaction networks for each community. By sliding the time window by one day, we create two consecutive networks that overlap significantly. In this way, we can capture subtle structural changes resulting from daily added/removed interactions. We calculate the different structural properties of these networks and analyze how they change over 180 days.

\subsection{Clustering}

There are many local and global measures of network properties \cite{boccaletti2006complex}. These measures are not independent. However, it was shown that the degree distribution, degree-degree correlations, and clustering coefficient are sufficient to fully describe most complex networks, including social networks \cite{orsini2015quantifying}. Furthermore, research on the dynamics of social group growth shows that links between persons' friends who are members of a social group increase the probability that that person will join that social group \cite{backstrom2006group}. Successful social diffusion typically occurs in networks with a high value of the clustering coefficient \cite{centola2007cascade}. These results suggest that higher local cohesion should be a characteristic of sustainable communities. 

The clustering coefficient of a node quantifies the average connectivity between its neighbors and the cohesion of its neighborhood \cite{boccaletti2006complex}.

It is a probability that two neighbours of a node $i$ are also neighbours, and is calculated using the following formula:
\begin{equation}
    c_{i}=\frac{e_{i}}{\frac{1}{2}k_{i}(k_{1}-1)} \ .
    \label{eq:clust}
\end{equation}
Here $e_{i}$ is the number of links between the neighbours of the node $i$, while $\frac{1}{2}k_{i}(k_{i}-1)$ is the maximum possible number of links determined by the degree of the node $k_{i}$. The clustering coefficient of the network $C$ is the value of the clustering averaged over all nodes. We investigate how the clustering coefficient in an SE community changes over time by calculating its value for all network snapshots. We compare the clustering behavior for active and closed communities on the same topic to better understand the evolution of cohesion of these communities.\\

\subsection{Core-periphery structure}

Real networks, including social networks, have a distinct mesoscopic structure \cite{fortunato2010community, gallagher2020clarified}. The Mesoscopic structure is manifested either through community structure or core-periphery structure. Networks with a community structure consist of a certain number of groups of nodes that are densely connected, with sparse connections between groups. Networks with core-periphery structures consist of two groups of nodes, with higher edge density within one group, core, and between groups. However, low edge density in the second group, periphery \cite{gallagher2020clarified}. Research on user interaction dynamics in SE communities shows that there is a small group of highly active members who have frequent interactions with casual or low active members \cite{santos2019activity, santos2019self}. These results indicate that we should expect a core-periphery structure in SE communities. Classification of nodes into one of these two groups provides information about their functional and dynamical roles in the network. 

To investigate the core-periphery structure of SE communities and how it evolves through time, we analyze the core-periphery structure of every network snapshot. For this purpose, we use the Stochastic Block Model (SBM) adapted for the inference of the core-periphery of the network structure \cite{gallagher2020clarified}. \\

SBM is a model where each node belongs to one group in the given network $G$. For the core-periphery structure, the number of blocks is two. Thus, the elements of the vector $\theta_i$ are $1$ if the node $i$ belongs to the core or $2$ for the periphery. The block connectivity matrix $\{{\vectorbold{p}}\}_{2x2}$ specifies the probability $p_{rs}$ that nodes from group $r$ are connected to nodes in group $s$, where $r,s\in\{1,2\}$.

The SBM model seeks the most probable model that can reproduce a given network G. The probability of having model parameters $\vectorbold{\theta}$, $\vectorbold{p}$ given network $G$ is proportional to likelihood of generating network $G$, $P(G | \vectorbold{\theta} , \vectorbold{p})$, prior on SBM matrix $P(\vectorbold{p})$ and prior on block assignments $P(\vectorbold{\theta})$:

\begin{equation}
   P(\vectorbold{\theta}, \vectorbold{p}| G) = P(G | \vectorbold{p} , \vectorbold{\theta}) P(\vectorbold{p}) P(\vectorbold{\theta}) \ ,
\end{equation}

The likelihood of generating a network $G$ is defined as:

\begin{equation}
  P(G | \vectorbold{\theta} , \vectorbold{p}) = \prod_{i<j} p_{r_is_j}^{A_{ij}}(1-p_{r_is_j})^{1-A_{ij}} \ ,
\end{equation}

where the adjacency matrix element $A_{ij}$ is equal to  $1$ whenever nodes $i$ and $j$ are connected and it is $0$ otherwise. 

Prior on $\bf{p}$ is uniform distribution over all block matrices whose elements satisfy constrain for core-periphery structure $0<p_{22}<p_{12}<p{11}<1$.
Prior on $\vectorbold{\theta} $ consists of three parts: probability of having $2$ blocks;  given the number of blocks, probability $P(n|2)$ of having groups of sizes $\{n_1, n_2\}$ and probability $P(\vectorbold{\theta}|n)$ of having particular assignments of nodes to blocks. 

To fit the model, we follow the procedure set by the authors of Ref. \cite{gallagher2020clarified} and use the Metropolis-within-Gibbs algorithm. For each 30 days snapshot network, we run 50 iterations and choose the model parameters $\theta$ and $p$ according to the minimum description length (MDL). MDL does not change much among inferred core-periphery structures, see Fig. A1 in SI, while looking into the Adjusted Rand Index (ARI), we can notice that difference exists. Still, the ARI between pair-wise compared partitions is significant ($ARI >0.9$), indicating the stability of the inferred structures. The definition and detailed descriptions of MDL and ARI are given in the SI.

\subsection{Dynamic reputation model}

Any dynamical trust or reputation model has to take into account distinct social and psychological attributes of these phenomena in order to estimate the value of any given trust metric \cite{duma2005dynamic}. First, the dynamics of trust are asymmetric, meaning that trust is easier to lose than to gain. As part of asymmetric dynamics, to make trust easier to lose, the trust metric has to be sensitive to new experiences, recent activity, or the absence of the user's activity while still maintaining the non-trivial influence of old behavior. The impact of new experiences must be independent of the total number of recorded or accumulated past interactions, making high levels of trust easy to lose. 
Finally, the trust metric must detect and penalize behavior that deviates from community norms. \\

We estimate the dynamic reputation of SE users using the Dynamic Interaction Based Reputation Model (DIBRM) \cite{melnikovDynamicInteractionBasedReputation2018}. This model is based on the idea of dynamic reputation, which means that the reputation of users within the community changes continuously over time: it should rapidly decrease when there is no registered activity from the specific user in the community, reputation decay, and it should grow when frequent, constant interactions and contributions to the community are detected. The highest growth in users' reputations is found through bursts of activity followed by a short period of inactivity. \\

Our model implementation does not distinguish between positive and negative interactions in SE communities. Therefore, we treat any interaction in the community, posting a question, answer, or comment, as a potentially valuable contribution. The evaluation criteria for SE websites that go through beta testing described in SI do not distinguish between positive and negative interactions.
The percentage of negative interactions in the communities we investigated was below 5\%, see Table A2 in SI. Filtering positive interactions would also require filtering out comments because the community does not rate them. That would eliminate a large portion of direct interactions between the community users, which is essential for estimating their reputation. The only negative aspect of behavior in our model is the absence of valuable contributions - the user's inactivity. This behavior can be seen as a deviation from community norms as we look at new communities in the early stages of development, where constant contributions are crucial to community growth and survival. \\

In DIBRM, the reputation value for each user of the community is estimated by combining three different factors: 1) \textit{reputation growth} - the cumulative factor that represents the importance of users' activities; 2) \textit{reputation decay} - the forgetting factor that represents the continuous decrease in reputation due to inactivity; 3) \textit{the activity period factor} - measuring the length of the period in which the change of reputation occurred. In the case of SE communities, the forgetting factor has a literal meaning, as we can assume that active users forget users' past contributions as their attention is captured by more recent content. \\

The reputation dynamics revolve around the varying influence of past and recent behavior. Thus, DIBRM has two components: \textit{cumulative factor} - estimating the contribution of the most recent activities to the overall reputation of the user; \textit{forgetting factor} - estimating the weight of past behavior. Estimating the value of recent behavior starts with the definition of the parameter storing the basic value of a single interaction $I_{b_{n}}$. The cumulative factor $I_{c_{n}}$ then captures the additive effect of successive recent interactions. The reputational contribution $I_n$ of the most recent interaction $n$ of any given user is estimated in the following way:

\begin{equation}\label{eq:ibn}
    I_n = I_{b_{n}} + I_{c_{n}} = I_{b_{n}} (1+  \alpha  (1-\frac{1}{A_{n}+1})) \ .
\end{equation}

Here, $\alpha$ is the weight of the cumulative part, and $A_{n}$ is the number of sequential activities. If there is no interaction at $t_n$, this part of interactions has a value of 0. An essential property of this component of dynamic reputation is the notion of sequential activities. Two subsequent interactions made by a user are considered sequential if the time between those two activities is less than or equal to the time parameter $t_{a}$ that represents the time window of interaction. This time window represents the maximum time spent by the user to make a meaningful contribution, post a question or answer or leave a comment,  

\begin{equation}\label{eq:deltan}
\Delta_{n}=\frac{t_{n}-t_{n-1}}{t_{a}} \ .
\end{equation}

If $\Delta_{n} < 1$ is less than one, the number of sequential activities $A_{n}$ will increase by one, which means that the user continues to communicate frequently. On the other hand, large values $\Delta_{n}$ significantly increase the effect of the forgetting factor. This factor plays a vital role in updating the total dynamic reputation of a user at each time step, after every recorded interaction:

\begin{equation}\label{eq:tn}
    T_{n}=T_{n-1} \beta^{\Delta_{n}}+I_{n} \ .
\end{equation}

Here, $\beta$ is the forgetting factor. In our model implementation, the trust is updated each day for every user regardless of their activity status. Therefore, the decay itself is a combination of $\beta$ and $\Delta_n$: the more days pass without recorded interaction from a specific user, the more their reputation decays. Lower values of $\beta$ lead to faster trust decay, as shown in Fig. A2 in the SI.
For this work, we select the following values of these parameters: 1) we set basic reputation contribution $I_{bn}=1$, meaning that each activity contributes 1 to the dynamical reputation; 2) for the cumulative factor $\alpha$ we choose the value $2$ and put higher weight on recent successive interactions; 3) forgetting factor $\beta$ we select the value $0.96$; 4) the value of $t_{a}=2$. By setting $\alpha>1$ we enable faster growth of trust due to a large number of subsequent interactions; see Fig. A2 in SI. Furthermore, by setting the value of $\beta<1.0$, we increase the penalty for long inactivity periods; see Fig. A2 in SI. We discuss the selection of model parameters and their consequences in detail in section \textit{Discussion}. The selected values of parameters are used to measure the dynamical reputation of users in all four pair SE communities. Given these parameter values, the minimal reputation of the user immediately after having made an interaction in the SE community is $1$. This reputation will decay below $1$ if the user does not perform another interaction within the one-day window. Users with a reputation below the value $1$ are considered inactive and \textit{invisible} in the community; that is, their past contributions at that time are unlikely to impact other users.  

\section{Results \label{sec:res}}

\subsection{Clustering and core-periphery structure of knowledge-sharing networks}

\begin{figure}[ht!]
    \centering

    \includegraphics[width=\linewidth]{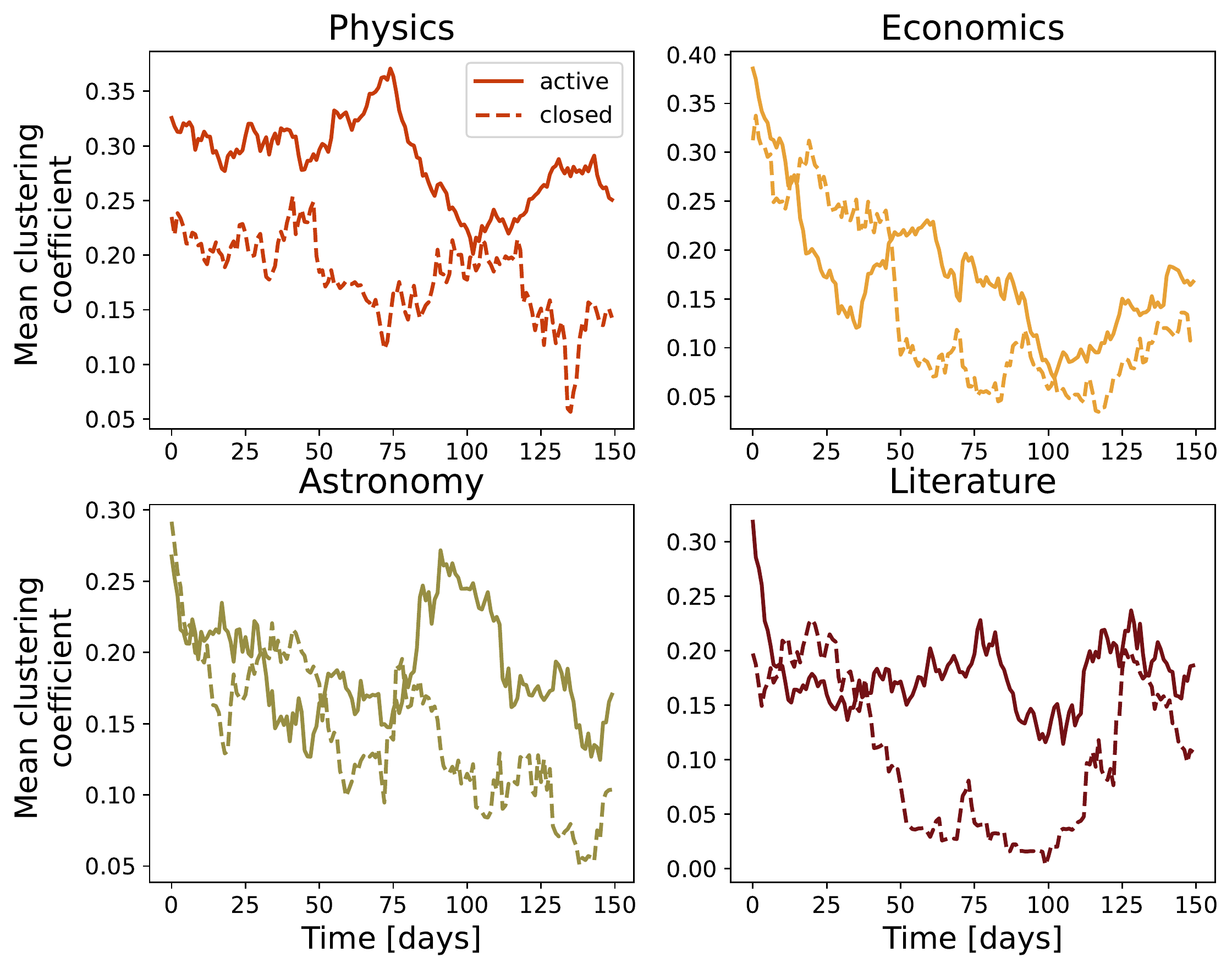}
    \caption{\textbf{Mean clustering coefficient of 30 days sub-networks for four pairs of Stack Exchange websites: }Astronomy, Literature, Economics, and Physics. Solid lines -- active sites; dashed lines -- closed sites. }
    \label{fig:clustering}
\end{figure}

We first analyze the structural properties of SE communities and examine the difference between active and closed ones. We calculate the mean clustering coefficient for 30-day window networks and examine how it changes over time. Figure \ref{fig:clustering} shows the evolution of the mean clustering coefficient for the eight communities. All communities that are still active are clustered, with a relatively high value of clustering coefficient, with Physics, the only launched community, having the highest value of clustering coefficient during the first 180 days.
During the larger part of the observed period, an active community's clustering coefficient is higher than the clustering coefficient of its closed pair. For pairs where active communities are still in the beta phase, the closed communities have a higher value of the mean clustering coefficient in the first 75 days. After this period, active communities have higher values of the clustering coefficient. These results suggest that all communities have relatively high local cohesiveness and that lower values of the clustering coefficient in the later phase of community life may indicate its decline. 

\begin{figure*}[ht!]

    \centering
    \includegraphics[width=\linewidth]{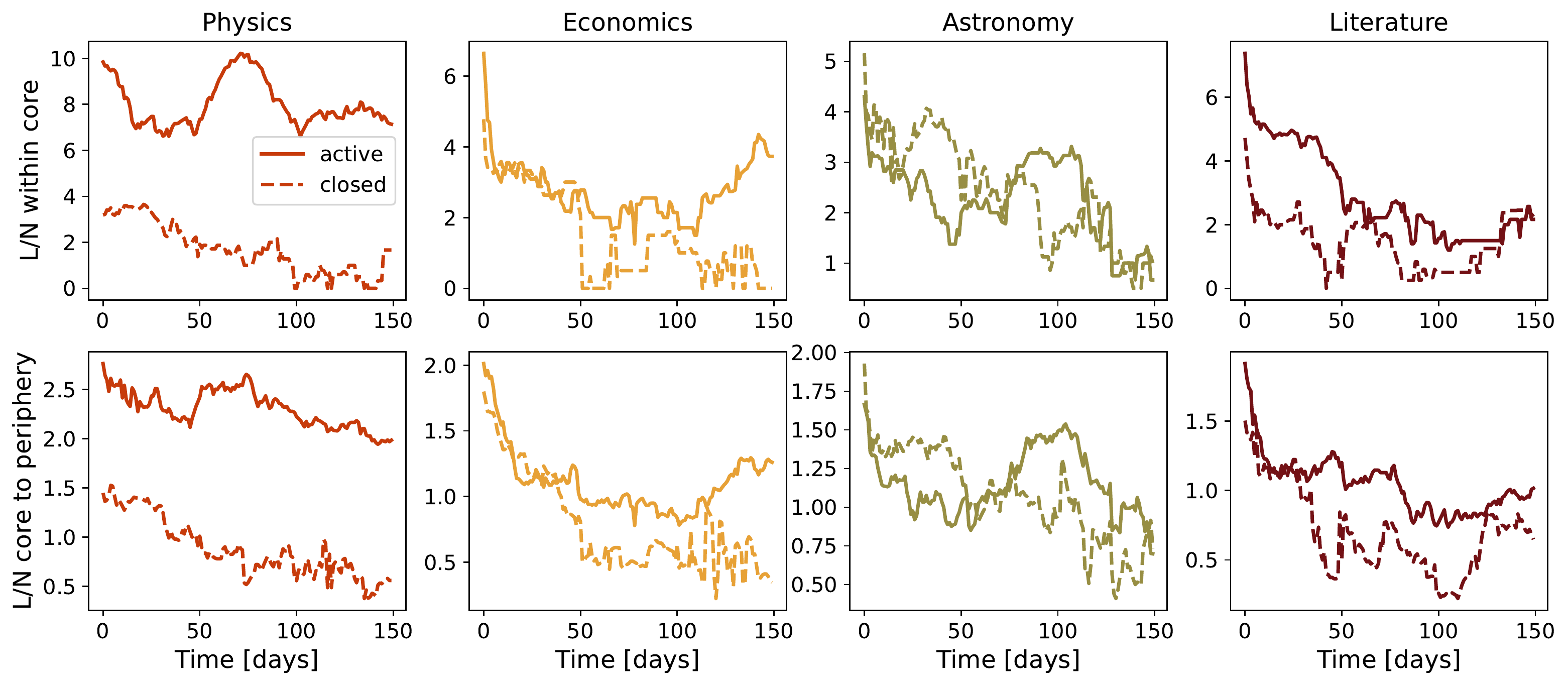}
    \caption{\textbf{Connectivity among users within the core and between core and periphery.} Links per node in core - top panel and links per node between core and periphery - bottom panel for the four pairs of Stack Exchange websites: Astronomy, Literature, Economics, and Physics. Solid lines -- active sites; dashed lines -- closed sites.}
    \label{fig:LPN}

\end{figure*}

Furthermore, we examine the core-periphery structure of these communities and their evolution. Specifically, we are interested in the evolution of connectivity in the core. Figure \ref{fig:LPN} shows the change in the number of links between nodes, averaged of the core nodes, $\frac{L_{c}}{N_{c}}$ over time. $\frac{2L_{c}}{N_{c}}$ is the average degree of the node in the core and, thus, $\frac{L_{c}}{N_{c}}$ is the half of the average degree. Again, the Physics community has a much higher value of this quantity than Theoretical physics during the observed period, indicating higher connectivity between core members. Higher connectivity between core members in the active community is also characteristic of Literature. However, this quantity has the same value for active and closed communities at the end of the observation period. The differences between active and closed communities are not that prominent for Economics and Astronomy, see Fig. \ref{fig:LPN}. Active and closed Economics communities have similar connectivity in the core during the first 50 days. After this period, the connectivity in the core of the active community is twice as large as in the closed community, and the difference grows at the end of the observation period. The connectivity in the core of the closed Astronomy community is higher than the connectivity in the core of the active community during the first 50 days. However, as time progresses, this difference changes in favor of the active community, while this difference disappears at the end of the observation period.

The difference between active and closed communities is observed compared to the average number of core-periphery edges per network node. The connectivity between core and periphery is higher for the active communities than for the closed ones, see Fig. \ref{fig:LPN}, which is very obvious if we compare Physics and Theoretical physics communities. Moreover, the Physics community has the highest connectivity compared to all other communities. Active Literature and Economics communities have the same core-periphery connectivity as their closed counterpart. The core of the active Astronomy community has weaker connections with the periphery than the closed community during the first 50 days, see Fig. \ref{fig:LPN}. 

Our motivation to examine the core-periphery structure comes from reference \cite{santos2019self}. The authors have selected $10\%$ of the most active users and examined their mutual connectivity and connectivity with the remaining users. The split of $10\%$ to $90\%$ users according to their activity may appear arbitrary. The core-periphery provides a more consistent network division based on its structure. However, the connectivity patterns between popular-popular and popular-casual users, shown in Fig. A3 in SI, are similar to one observed for core-periphery in Fig. \ref{fig:LPN}.

On average, the cores of active communities have a higher number of nodes in the core than closed communities, Fig. A4. However, the relative size of the core compared to the size of the whole network, Fig. A4 in SI, is similar for active and closed communities. The size of the core fluctuates over time for active and closed communities. The core membership also changes over time. This core membership is changing more for the closed communities. We quantify this by calculating the Jaccard index between the cores of the subnetworks at the moment $t_{i}$ and $t_{j}$. Figure A5 in SI shows the value of the Jaccard index between any pair of the 150 subnetworks. The highest value of the Jaccard index is around the diagonal and has a value close to 1. The compared subnetworks are for consecutive days and have a similar structure. The value of the Jaccard index decreases with the number of days between two subnetworks $|t_{i}-t_{j}|$ faster in closed communities; see Fig. A6 in SI. This difference is the most prominent for the literature communities, while this difference is practically non-existent for Astronomy. The relatively high value of overlap between cores of distant subnetworks for active communities further confirms that the core is more stable in these communities that in their closed counterparts. 

\subsection{Dynamic reputation of users within the network of interactions}

\begin{figure*}
    \centering
    \includegraphics[width=\linewidth]{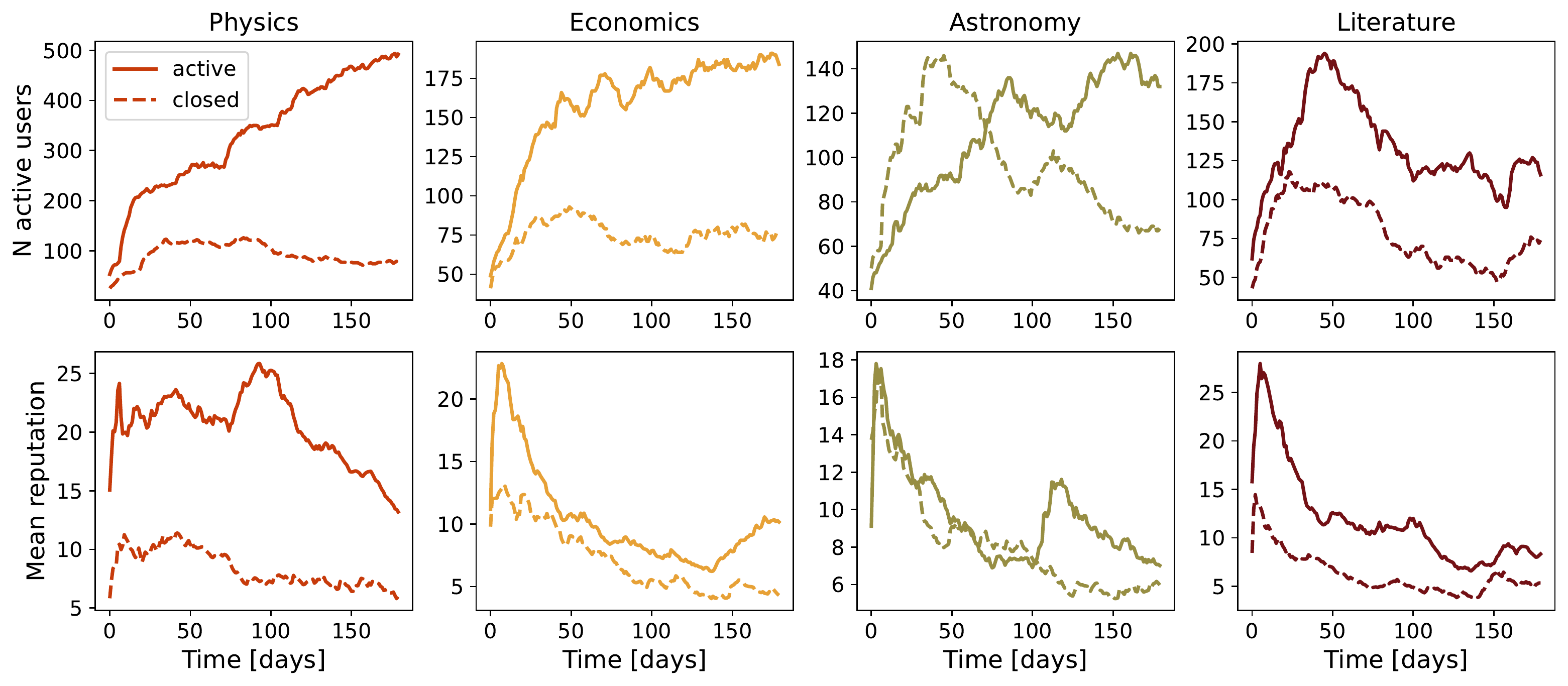}
    \caption{\textbf{Active users within SE communities and their mean dynamic reputation.} The number of active users (users with a reputation higher than 1) - top panel, and mean Dynamic Reputation within active users -- bottom panel for the four pairs of Stack Exchange websites: Astronomy, Literature, Economics, and Physics. Solid lines -- active sites; dashed lines - closed sites.}
    \label{fig:dr6panel}
\end{figure*}

\begin{figure}
    \centering
    \includegraphics[width=\linewidth]{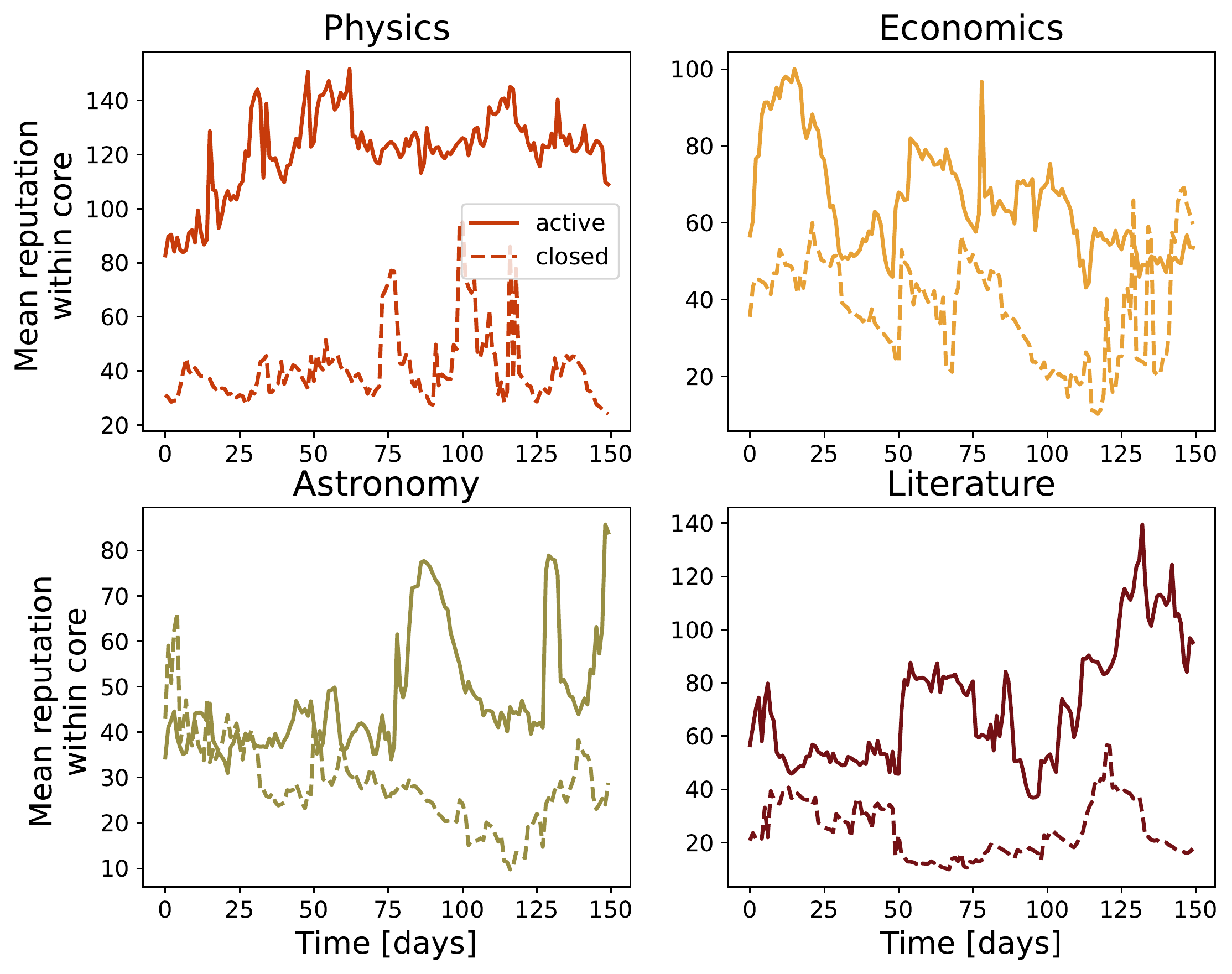}
    \caption{Mean Dynamical reputation within the core for four pairs of Stack Exchange websites: Astronomy, Literature, Economics, and Physics. Solid lines -- active sites; dashed lines -- closed sites.}
    \label{fig:dr_core}
\end{figure}

To explore the differences between active and closed communities, we focus on dynamical reputation, our proxy for collective trust in these communities. 
The number of active users (top panel) and the mean user reputation (bottom panel) for different SE communities are shown in Figure \ref{fig:dr6panel}. Except in the case of astronomy, closed communities generated less engaged users from the start and the number of active users saturated at lower values. In the case of astronomy, the closed community started with a faster-increasing number of active users. However, within the first two months, their number dropped, while the second time around, the community started slower but kept engaging more users. Only in the still active physics community is the number of active users an increasing function over the whole 180 day period we have observed. Panels in the bottom show mean reputation among active users, and we see that most of the time, it was higher in the still active communities than in the closed ones. The Physics community kept these mean values more stably at higher levels, whereas in other communities, we note that the initial high mean reputation decays faster. Astronomy is an exciting exception again, where we see a second sudden increase in mean user reputation, which signals an increase in user activity.

In addition, we investigate whether and how the core-periphery structure is related to collective trust in the network. Figure \ref{fig:dr_core} shows the mean dynamical reputation in the core of active and closed communities and its evolution during the observation period. There are apparent differences between active and closed communities regarding dynamical reputation. The mean dynamical reputation of core users is always higher in active communities than in closed. The most significant difference is observed between the Physics and Theoretical Physics communities. The difference between active communities, which are still in the beta phase, and their closed counterparts is not as prominent. However, the active communities have a higher mean dynamical reputation, especially in the later phase of the observation period. The only difference in the pattern is observed for astronomy communities at the early stage of their life. The closed community has a higher value of dynamic reputation than the active community. This observation is in line with similar patterns in the evolution of mean clustering, core-periphery structure, and mean reputation. 

By definition, the core consists of very active individuals. Thus we expect a higher total dynamical reputation of users in the core than the total reputation of users belonging to subnetworks periphery. Figure A7 shows the ratio between the total reputation of the core and periphery for closed and active communities and their evolution. The ratio between the total reputation of core and periphery in Physics is always higher than in the Theoretical physics community. A similar pattern can be observed for literature communities, although the difference is not as prominent as in the case of physics. The ratio of total dynamical reputation between core and periphery was higher in the closed Economics community during the early days of its existence. However, this ratio becomes higher for active communities in the later stage of their lives. Communities around the astronomy topic deviate from this pattern, which shows the specificity of these two communities. 

To complete the description of the evolution of dynamic reputation, we examine the evolution of the Gini index of dynamical reputation among the active members of SE sites, shown in Fig. A8 in SI. 
Both closed and active communities have high values of the Gini index, indicating that the dynamic reputation is distributed unequally among users. Notably, all communities have the highest Gini index at the start, signaling that the inequality in users' activity at the start, and thus their dynamic reputation is the highest. After this initial peak, the Gini index decreases, but it persists at higher levels in communities that are still active than in the closed ones, except in the case of the Astronomy community.
In this case, the active community had a higher Gini index until just before the observation period, when the Gini coefficient increased in the closed community. 

Figure A9 in SI shows the evolution of the assortativity coefficient for users' dynamical reputation. The observed networks are disassortative during the most significant part of 180 days period. Users with high dynamical reputations tend to connect with users with a low value of dynamical reputation in all eight communities. We also compare the degree and betweenness centrality of the users and their dynamical reputation by calculating the correlation coefficient between these measures for each sliding window, see Fig. A10 and detailed explanation in SI. The correlation between these centrality measures and dynamical reputation is very high. In active communities on physics, economics, and literature topics, the correlation between centrality measures and users' reputation is exceptionally high, above $0.85$, and does not fluctuate much during the observation period. There is a clear difference between active and closed communities for these three pairs. The astronomy pair deviates from this pattern for the first $100$ days. After this period, the pattern is similar to one observed for the other three pairs of communities. The results reveal that degree and betweenness centrality are correlated more with a reputation in active than in closed communities.

\section{Discussion and conclusions \label{sec:cons}}

In this work, we have explored whether the structure and dynamics of social interactions determine the sustainability of knowledge-sharing communities. We have adopted a model of dynamical reputation to measure the collective trust of members and analyzed its dynamics. For this purpose, we use the data from the SE platform of knowledge-sharing communities where members ask and answer questions on focused topics. We selected four pairs of active and closed communities on the same or similar topic. Specifically, two topics are from the STEM field, physics, and astronomy, and two from social and humanities topics, economics and literature.

We have examined the evolution of the average clustering coefficient in closed and active SE communities. Our results show that active communities have higher clustering coefficient values in the later phase of community life. In the early phase of communities' lives, the clear difference between active and closed communities is observed only for the physics topic; see Fig. \ref{fig:clustering}. The high value of the clustering coefficient observed for the active Physics community, the only considered community that has graduated, suggests that communities with high local cohesiveness are successful and mature faster than others. 

The core in active communities is strongly connected with the periphery than in closed communities, indicating that active members engage more often with occasionally active members; see Fig. \ref{fig:LPN}. These results suggest that active communities are more inclusive than closed ones. Furthermore, our analysis shows that average connectivity between core members is not as crucial to community sustainability as expected. Although active communities on physics and economics exhibit much higher connectivity in the core than their closed counterparts, this is not true for communities focused on astronomy and literature. However, our results show that a member's lifetime in the core is longer for active communities, indicating a more stable core in active communities. 

Analysis of the evolution of the core-periphery and its connectivity patterns suggests a higher trust between active and sporadically active members. To further explore this, we have adapted the dynamical reputation model \cite{melnikovDynamicInteractionBasedReputation2018}, which allowed us to follow the evolution of trust of each member.

The total dynamical reputation of core members during their first 180 days was higher for active communities than for their closed counterparts. While relative core size is less than $40\%$, Fig. A4 in SI, the ratio between the total reputation of nodes in the core and ones in the periphery is consistently above 0.5, indicating that the average reputation of members in the core is higher than the reputation of the node in the periphery. The ratio between the total reputation of core and periphery nodes has a higher value in the active community of Physics, Literature, and Economics active community. For most of the 180 days, this ratio has a value higher than one. The Astronomy communities are outliers, but the core members have a higher total reputation than members on the periphery, even for these two communities. Our results imply that the most trusted members in the community are the core members, who also generate more trust in active communities. They have a higher reputation generated through interactions with both core and nodes in the periphery, Fig. \ref{fig:dr_core}. Furthermore, the overall levels of trust are higher in active communities, which is reflected in the fact that the mean user reputation is higher in these communities; see Fig. \ref{fig:dr6panel}.

The choice of the topics and selection of SE communities of a various number of users, question, answer and comments, see Tab. \ref{tab:site-info}, guarantees, up to a certain extent, the generality of our results. However, there are certain limitations to the generalizability of our findings. While SE communities provide very detailed data that enable the study of the structure and dynamics of knowledge-sharing communities, we must not ignore the fact that they have some properties that make them specific. 

SE communities are about specific topics; they mostly bring together people who are passionate about or are experts in a specific field. These communities attract people from the general population. Since we were interested in excluding the factor of the topic in our research, we studied and compared successful and unsuccessful communities on the same topic. In the SE network, these pairs of communities are pretty rare, which has substantially limited our sample size, leaving the possibility for the occurrence of outliers that do not follow our general conclusions. 

Finally, there are many ways to measure collective trust and reputation in online social communities. We have selected the dynamical reputation model because it was developed to measure reputation in SE communities. Furthermore, the model allowed us to study the evolution of trust in communities. However, the model requires fine-tuning of its parameters and does not distinguish positive from negative interactions. We have selected our parameters to replicate the activity of the SE communities in the time window of $\tau=30$ days. Our analysis shows that while the choice of the sliding window, $\tau$, may seem arbitrary, the different values do not influence the general conclusions; see Fig. A11 in SI. The interactions in SE communities are mostly not emotional, and thus, the model is suitable for measuring collective trust in these communities. However, the interaction in other knowledge-sharing communities can be much more emotional, and therefore the dynamical reputation model needs to be adapted to measure reputation in these communities. \\

Our results show that the trustworthiness of core members thus represents one of the essential parameters for determining community sustainability. Sustainable communities have a core of trustworthy members. The core of sustainable communities is more densely connected, and its connectivity with the periphery is more significant than in closed communities. The observed feature is especially prominent in the Physics community, which is the only active community considered to be mature. As we stated, active communities on topic of astronomy, economics and literature were in the beta phase. However, since of December 2021 \footnote{\url{
https://stackoverflow.blog/2021/12/16/congratulations-are-in-order-these-sites-are-leaving-beta/}} , these communities graduated. The core of sustainable communities exhibits higher degrees of stability during their first 180 days. Sustainable communities have higher local cohesiveness, which is reflected in the relatively high value of the clustering coefficient. Our results show that these conclusions hold for both STEM and humanities topics. However, we do not observe apparent differences between active and closed astronomy communities for some quantities. In the case of astronomy and sometimes economics, we find that closed communities had higher clustering coefficients and higher core-core and core-periphery connectivity during the early phase of community life. These observations suggest that the properties of the network during the early phase of the community's existence may lead to wrong conclusions about its sustainability. Our results also imply that information about community sustainability is hidden in the evolution of different network and trust properties.      

\subsection{The choice of model parameters}

In this work, we used snapshots of the network of 30 days. This period corresponds to the average month, and it is common in the analyses of the structure and dynamics of social networks \cite{saramaki2015seconds,krings2012effects,barrat2021temporal}. Still, there is no well-specified procedure for choosing the time window. Previous studies have shown that if $\tau$ is small, subnetworks become sparse, while for too large sliding windows, some important structural changes cannot be observed \cite{krings2012effects, arnold2021moving}. Thus, we have analysed how the time window choice influences our results. Figure A11 in SI shows how considered network properties and dynamical reputation depend on the time window size for active and closed communities on the astronomy. We observe that fluctuations of all measures are more pronounced for a time window of 10 days than for 30 and 60 days. However, we find that while the structural properties of networks evolve at different rates over varied time windows, the trends remain very similar. The observed qualitative difference between closed and active communities is independent of the time window size, especially when comparing the 30 and 60 day windows. The 30-day time window ensures enough interaction, even for closed communities, while the number of observation points remains relatively high. For these reasons, we choose a sliding window of 30 days.\\

The initial purpose of DIBRM was to replicate the dynamics of SE’s official reputation metric \cite{melnikovDynamicInteractionBasedReputation2018,yashkina2020}. As SE’s reputation is harder to be lost than gain, in previous studies \cite{yashkina2020} the official SE reputation is obtained with $t_a =2, \beta = 1, \alpha = 1.4$, which means that there is no active forgetting factor. Our application is oriented towards estimating a reputation metric concerning fundamental properties of social trust, i.e. reputation decreases with members' inactivity,  so we opted for a different set of parameter values.

For the basic reputation contribution of a single interaction, we selected $I_{bn} = 1$, and at the same time, this is the threshold value of an active user. This value is intuitive as every interaction has initial contribution of +1 to the user's reputation, although the previous works have used values of +2 and +4. Following the previous work and after examining the median/average time between subsequent interactions of the same user, we selected $t_a = 1$, which also means that the reputation in our model will be updated every day during the time window of the analysis, regardless of whether the user is active or not. 

The combination of parameters $\alpha$ and $\beta$ can significantly influence the dynamic of the single user reputation, as shown in Fig A2. We show that higher values for parameter $\alpha=2$, highlight the burst of user activity and frequent interaction. On the other hand, the parameter beta is the forgeting factor, which at the same time determines the weight of past interactions and the reputational punishment due to user inactivity. Here, we need to select the  parameter $\beta$ value, so we include forgetting due to inactivity but do not penalize it too much. In Fig. A2, we show how different values of parameter $\beta$ influence the time needed for a user's reputation to fall on value $I_{n}=1$ due to the user's inactivity and value of dynamical reputation at the moment of the last activity. The higher the value of the parameter $\beta$ and the initial dynamical reputation of the users, the longer it takes for the user's reputation to fall to the baseline value. For parameters $\beta=0.9$ and $I_{n}=5$, the user's reputation drops to value $I_{n}=1$ after less than 20 days, while this time is doubled for $\beta=0.96$. We see that for higher values of the parameter $\beta$, the time it takes for $I_{n}$ to drop to $1$ becomes longer and that the initial value of the reputation becomes less important.\\

We estimated the difference between the number of users who had at least one activity in the 30-day window and the number of users with a reputation greater than $1$ during the same period for different parameter $\beta$ values. We calculated the root mean square error (RMSE) between the time series of the number of active users for $\tau=30$ and different values of $\beta$ parameters; see Fig. A12 in SI. The minimal difference between these two variables is for $\beta$ between $0.94$ and $0.96$ for both active and closed communities. Since we want to compare communities, we select $\beta = 0.96$. Our analysis reveals that the reputational decay parameter $\beta$ set at $0.96$ does not reduce the number of active users (based on their dynamic reputation) below the actual number of users who have been active (interacted with the community) in the time window of 30 days; see Fig. A13 in SI. Furthermore, we examine and compare the trends of two types of time series: 1) time series of active users, according to dynamical reputation; 2) time series of permanent users, users who were active in a given sliding window and continued to be active in the next one. Figure A14 in SI shows that while the absolute number of users differs in these time series, they follow similar trends for all communities.

\textbf{Declarations of interest:} None
\section{Acknowledgements}
A.A., A.V. and M.M.D. acknowledge funding provided by the Institute of Physics Belgrade, through the grant by the Ministry of Education, Science, and Technological Development of the Republic of Serbia. Numerical simulations were run on the PARADOX-IV supercomputing facility at the Scientific Computing Laboratory, National Center of Excellence for the Study of Complex Systems, Institute of Physics Belgrade.
\bibliographystyle{TRR}
\bibliography{ref.bib}

\renewcommand{\thefigure}{A\arabic{figure}}
\setcounter{figure}{0}

\renewcommand{\thetable}{A\arabic{table}}
\setcounter{table}{0}
\onecolumn
\section{Supplementary Information}

\subsection{Area51 criteria}

The Stack Exchange network has its criteria for the success of sites. They measure how many questions are answered, how many questions are posted per day, and how many answers are posted per question.  There are two measures: the number of avid users and the number of visits that are not easily interpreted from the data. The site is \textit{healthy} if it has 10 questions per day, 2.5 answers per question and more than $90\%$ of answered questions. For less than $80\%$ of answered questions, 5 questions per day and 1 question per answer site \textit{needs some work}. We calculated Stack Exchange statistics for Astronomy, Economics, Literature and Physics communities and results are presented in the table \ref{tab:se_c}. After 180 days, only active Physics is a healthy site, whereas other active sites are at least in two criteria labeled \textit{okay}. Closed sites mostly \textit{need some work}, the exception is closed astronomy with \textit{excellent} percent of answered questions and \textit{okay} answer ratio. 

\begin{table}[H]
\footnotesize\sf\centering
\caption{Community overview for first 180 days according to SE evaluation criteria  }
\label{tab:se_c}
\begin{tabular}{cc|c|c|c} 
\hline
Site                        & \multicolumn{1}{c}{Status}          & \multicolumn{1}{c}{Answered}  & \multicolumn{1}{c}{Questions per day} & Answer ratio   \\ 
\hline
\multirow{2}{*}{Physics}    & Closed                              & 83 \%                         & 1.93                                  & \underline{1.64}   \\
                            & Active                            & \textbf{93} \%                & \textbf{11.76}                        & \textbf{2.74}  \\ 
\hline
\multirow{2}{*}{Economics}  & Closed                              & 68 \%                         & 2.04                                  & \underline{1.25}   \\
                            & Active                                & \underline{84} \%                 & \underline{5.66}                          & \underline{1.37}   \\ 
\hline
\multirow{2}{*}{Astronomy}  & Closed                              & \textbf{95} \%                & 2.62                                  & \underline{2.02}   \\
                            & Active                                & \textbf{96} \%                & 3.57                                  & \underline{1.49}   \\ 
\hline
\multirow{2}{*}{Literature} & Closed                              & 79 \%                         & 1.77                                  & \underline{1.65}   \\
                            & Active                                & 74 \%                         & \underline{5.04}                          & \underline{1.10}   \\ 
\hline
Stack Exchange criteria     & \multicolumn{1}{c}{excellent}       & \multicolumn{1}{c}{$>$ 90 \%} & \multicolumn{1}{c}{$>$10}             & $>$ 2.5        \\
                            & \multicolumn{1}{c}{needs some work} & \multicolumn{1}{c}{$<$ 80 \%} & \multicolumn{1}{c}{$<5$}              & $<$ 1          \\
\hline
\end{tabular}
\end{table}

\subsection{Negative interactions}

The average percentage of negatively voted interactions is 3.2\% for questions and 3\% for answers. Percentages for questions and answers from each community are shown in table \ref{tab:negint}.  Comments cannot have a negative vote sum as they can only be upvoted.

\begin{table}[H]
\footnotesize\sf\centering
\caption{Percentage of negatively voted interactions}
\label{tab:negint}
\begin{tabular}{cc|c|c} 
\hline
Site                        & Status   & \multicolumn{1}{c}{Questions} & Answers  \\ 
\hline
\multirow{2}{*}{Physics}    & Active & 5\%                           & 4\%      \\
                            & Closed   & 1\%                           & 2\%      \\ 
\hline
\multirow{2}{*}{Economics}  & Active     & 4\%                           & 4\%      \\
                            & Closed   & 7\%                           & 4\%      \\ 
\hline
\multirow{2}{*}{Astronomy}  & Active     & 3\%                           & 3\%      \\
                            & Closed   & 2\%                           & 1\%      \\ 
\hline
\multirow{2}{*}{Literature} & Active     & 2\%                           & 5\%      \\
                            & Closed   & 2\%                           & 1\%      \\ 
\hline
\textbf{Average}            &          & 3.2\%                         & 3\%      \\
\hline
\end{tabular}
\end{table}

\vskip -5cm

\subsection{Robustness of core-periphery algorithm}
Consider the network $G(V, L)$, with a set of nodes $V$ and a set of links between them $L$. The stochastic community detection algorithms may converge to different configurations. To quantify the similarity between the obtained structures and the robustness of the algorithm, we run 50 iterations and calculate several similarity measures between pairwise partitions $C$ and $C^{'}$.

The core-periphery structure has two groups, so the confusion matrix \cite{labatut2012accuracy} can be defined as:

\begin{center}
    
\begin{tabular}{l|l|c|c|c} 

\multicolumn{2}{c}{}&\multicolumn{2}{c}{partition C}&\\ 

\cline{3-4} 
\multicolumn{2}{c|}{}&core&periphery&\multicolumn{1}{c}{}\\
\cline{2-4} 
partition & core & $n_{TP}$ & $n_{FN}$ & \\ 
\cline{2-4} $C^{'}$ & periphery & $n_{FP}$ & $n_{TN}$ & \\ 
\cline{2-4}
\end{tabular}
\end{center}

The diagonal elements correspond to the number of nodes found in the same class in both node configurations. The number of nodes in the core found in $C$ and $C^{'}$ is denoted as true positive $n_{TP}$, while the number of nodes in the periphery in $C$ and $C^{'}$ is denoted as true negative $n_{TN}$. The off-diagonal elements of the confusion matrix indicate the number of nodes differently classified. We can define the number of nodes found in the first configuration C in the core but in $C^{'}$ in the periphery as a false positive, $n_{FP}$, similarly the number of nodes found in the periphery in the partition $C$, and in the core in partition $C^{'}$ as a false positive, $n_{FP}$. 

From the confusion matrix, we can write the precision $P =n_{TP}/(n_{TP}+n_{FP})$ and recall $R=n_{TN}/(n_{TN}+n_{FN})$. These measures range from 0 to 1. The precision (recall) corresponds to the proportion of instances predicted to belong (not belong) to the considered class and which indeed do (do not) \cite{labatut2012accuracy}. \\~\\

The \textbf{F1 measure} is the harmonic mean of precision and recall \cite{labatut2012accuracy}:
\begin{equation}
    F_1 = 2\frac{P \cdot R}{P+R} = \frac{2n_{TP}}{2n_{TP}+n_{FN} + n_{FP}}
\end{equation}
It can be interpreted as a measure of overlap between true and estimated classes; it is 0 for no overlap to 1 if overlap is complete. \\~\\

The \textbf{Jaccard's} coefficient is the ratio of two classes' intersection to their union \cite{labatut2012accuracy}. It can also be expressed in terms of confusion matrix: 
\begin{equation}
    J =  \frac{C_{core} \cap C^{'}_{core}}{C_{core} \cup C^{'}_{core}} =   \frac{n_{TP}}{n_{TP}+n_{FP} + n_{FN}}
\end{equation}

\textbf{Normalized mutual information (NMI)} is similarity measure between to partitions $C$ and $C^{'}$  based on information theory \cite{danon2005comparing}:

\begin{equation}
    NMI(C, C^{'}) = \frac{MI(C, C^{'})}{(H(C)+H(C^{'})/2}
\end{equation}

where $MI$ is mutual information between sets $C$ and $C^{'}$, while $H(C)$ is entropy of given partition. The entropy is defined as $H(C) = - \sum_{i=1}^{|C|}P(i)log(P(i))$, where $P(i) = |U_i|/N$ is the probability that an object is randomly classified as $i$ (in this special case $i=0$, the node belongs to the core, or $i=1$, the node belongs to the periphery). The mutual information between sets $C$ and $C^{'}$ measures the probability that the randomly chosen node is a member of the same group in both partitions:
\begin{equation}
MI(C, C) = \sum_i\sum_j P(i, j) log(\frac{P(i,j)}{P(i)P^{'}(j)})    
\end{equation}
where $P(i, j)= |U_i \cap U_j|/N$

$NMI$ ranges from 0 when the partitions are independent to 1 if they are identical.   

\textbf{Adjusted rand index.} For the set of nodes $V$, with $n$ nodes, consider all possible combination of pairs $(v_i, v_j)$. We can select the number of the pairs where nodes belong to the same group in both partitions, $C$ and  $C^{'}$, denoted as $a$. Similarly, as $b$, we can define the number of pairs whose nodes belong to different groups in partitions. Then, unadjusted rand index \cite{santos2009use} is given as $RI = \frac{a+b}{\binom{n}{2}}$, where $\binom{n}{2}$ is number of all possible pairs. The RI between two randomly assigned partitions is not close to zero; for that reason, it is common to use the adjusted rand index \cite{hubert1985comparing} defined as:
\begin{equation}
    ARI = \frac{RI - E[RI]}{max(RI)- E[RI]}
\end{equation}
where $E[RI]$ is expected value of RI, and $max(RI)$ is maximum value of $RI$.

\begin{figure}[H]
    \centering
    \includegraphics[width=0.8\linewidth]{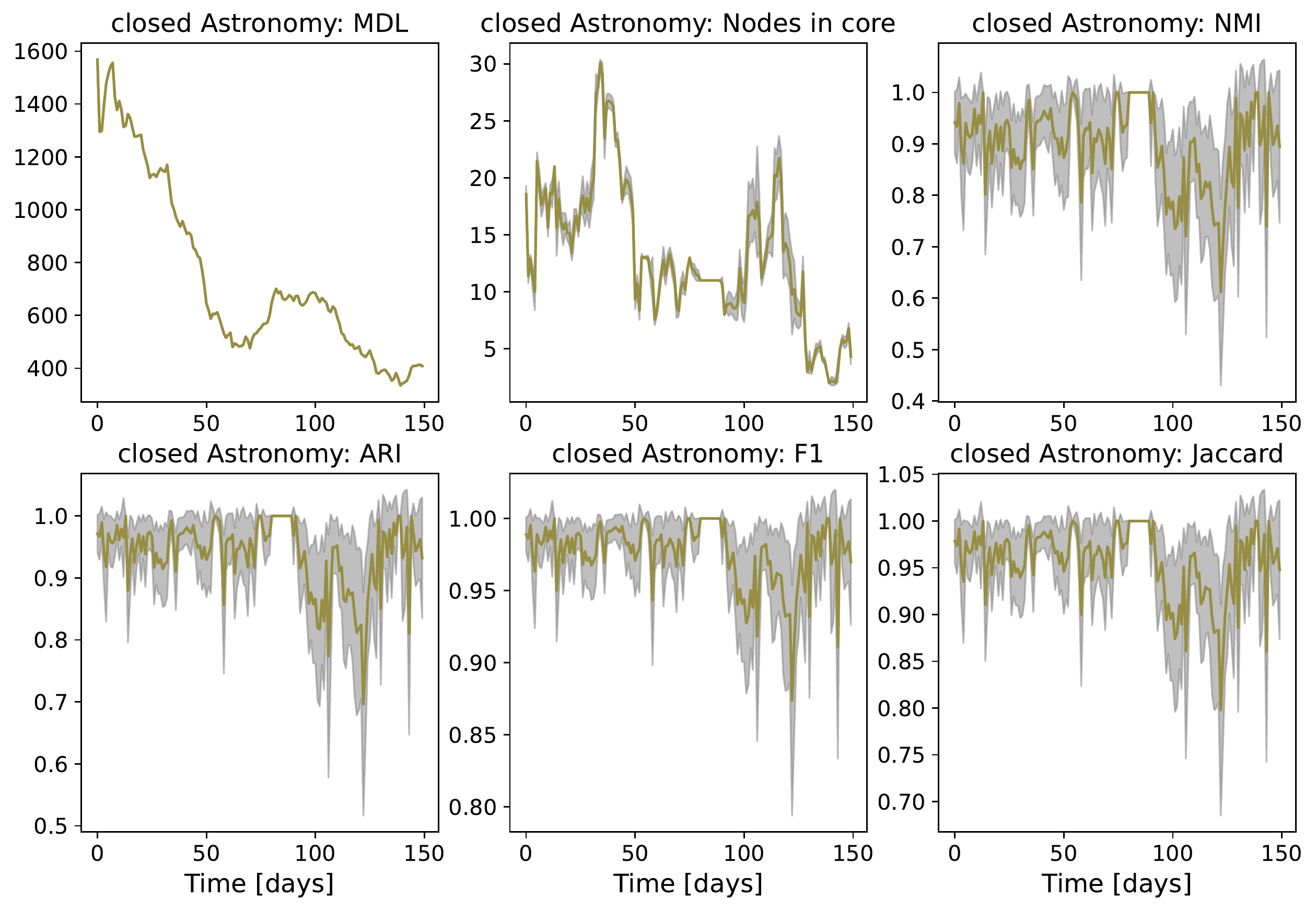}
    \caption{Minimum description length, number of nodes in core, normalized mutual information, adjusted rand index, F1 measure and Jaccard index, among 50 samples for 30-days sub-networks. Results are given for closed astronomy. }
    \label{fig:sample}
\end{figure}

For example, we show analysis of an inferred sample of core-periphery structures for 30 days closed Astronomy networks, Figure \ref{fig:sample}. We represent the mean minimum description length (MDL) and the mean number of nodes in the core with standard deviation. MDL does not change much between inferred core-periphery structures; the difference between obtained configurations is still notable in the number of nodes in the core.  To investigate similarity between obtained core-periphery configurations in the sample, we calculate several measures between pair-wise partitions such as normalized mutual information, adjusted rand index, F1 measure and Jaccard index. These measures are greater than 0.5 and, in most cases, greater than 0.9, indicating the stability of the inferred core-periphery structures.

\subsection{Dynamic reputation parameters}

Our implementation of the dynamic reputation model was based on $\beta = 0.96$. There are several reasons to select this value.

\begin{figure}[H]
    \centering
    \includegraphics[width=\linewidth]{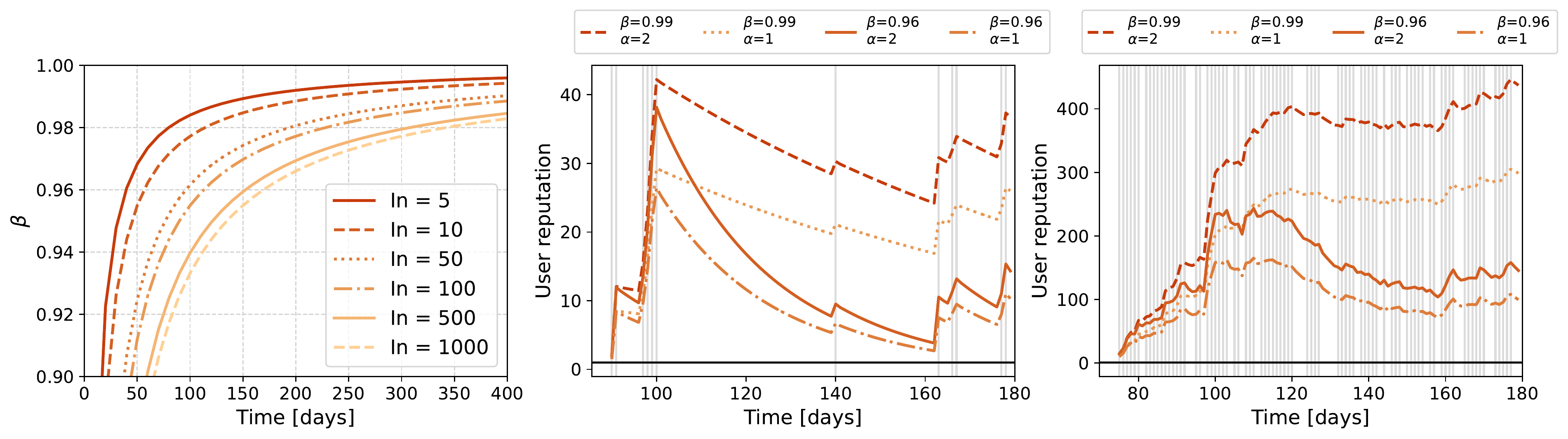}
    \caption{Dependence of parameter $\beta$ and number of days $\Delta$ needed for reputation $I_n$ to drop to baseline reputation $I_{n_0} = 1$. Dependence of parameter $\beta$ and number of days when reputation due inactivity decreases from $I_n$ to $I_0$ is given as  $\beta = (\frac{I_{n0}}{I_{n}})^{(1/\Delta)}$. The dynamic reputation of a single user, for different model parameters $\alpha$ and $\beta$. The middle graph shows the less active user, whose dynamic reputation tends to drop due to inactivity. On the other hand, frequently active user gains a significant reputation, and its decrease is slighter, as shown on the right graph.}
    \label{fig:betadelta}
\end{figure}

In the Dynamic reputation model, the $\beta$ parameter controls the strength of the forgetting factor of the model.  The value of this parameter should reflect the core feature of the reputation systems and make the reputation easier to lose. Due to the user's inactivity, any  reputation level will eventually decay to below 1. Dependence of time needed for reputation to drop below this level and the $\beta$ parameter, as well as reputation before inactivity is shown on Figure~\ref{fig:betadelta}. Here $I_n$ equals to the raw number of interactions in the community without forgetting or a cumulative factor at work.

For $\beta$ values below 0.96, the decay is fast, and within two to four months of inactivity, even high reputation values are reduced below the threshold. On the other hand, with values of $\beta$, the decay process is more differentiated, and the high reputation becomes harder to lose, surviving up to a year of inactivity. For $\beta$ equal to 0.96, it takes a month for the reputation based on 5 interactions to decay and around 5 months for the high reputation based on 500 or 1000 interactions to decay below the threshold.

Fig \ref{fig:betadelta}, as an example, shows the evolution of a single user reputation, one rarely active (middle graph) and frequently active (right graph). The combination of model parameters $\beta$ and $\alpha$ influence the dynamic of reputation. Note that with $\beta=0.96$, reputation can quickly drop close to the threshold, while with larger values of $\beta$, reputation stays larger even if a user is not active for more than a month. The most significant influence on the reputation of frequently active users has a parameter $\alpha$. The higher $\alpha$ highlights the activity burst, leading to higher reputation values. We decided to fix $\alpha=2$.

\subsection{Core-periphery structure of the interaction networks - core size and stability}

\begin{figure}[H]
    \centering
    \includegraphics[width=\linewidth]{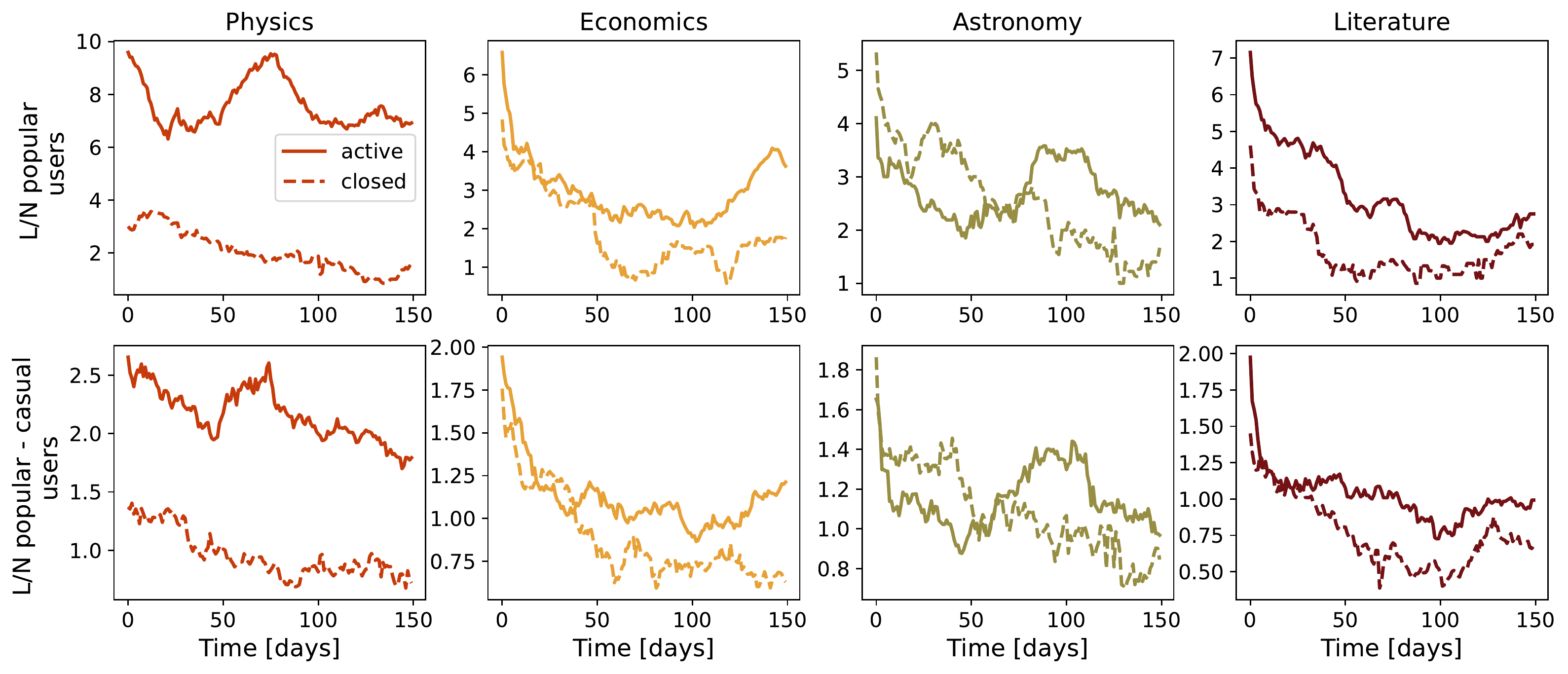}
    \caption{Links per node among popular users (top 10\% of users) and between popular and casual users.}
    \label{fig:pop_cas_users}
\end{figure}

There are commonly two types of users, in Q-A communities: popular and casual users. Popular users tend to generate the majority of interactions - they are likely to post more questions, take part in answering questions and engage in discussions through comments. For popular users, we consider $10\%$ of the most active users. We analyse interactions between popular and casual users and  among popular users in the sub-networks of 30 days [t+30). In both cases, the number of links per node for active sites is larger than in closed communities (Figure \ref{fig:pop_cas_users}).

Although this separation of users emphasizes differences between closed and active sites, it does not guarantee that all popular users are in the top 10. To solve this dilemma, we use the SBM (Stochastic Block Model) algorithm) to detect the core and the periphery of each 30 day sub-network. Such a split of users leads us to similar conclusions as before. (see figure \ref{fig:windows} - 2nd column).

\begin{figure}[H]
    \centering
    \includegraphics[width=0.9\linewidth]{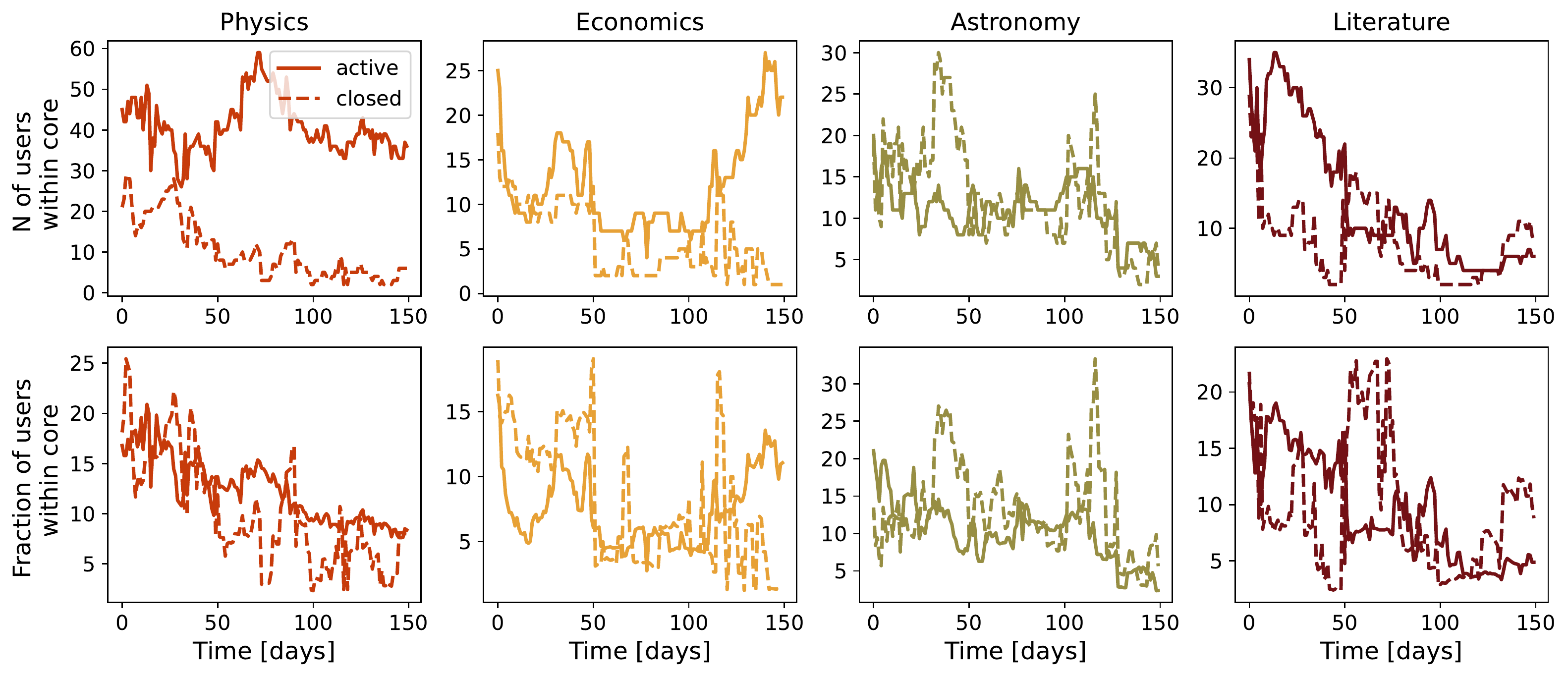}
    \caption{The size of the core (top) and fraction of users in core(bottom). Solid lines - active sites; dashed lines - closed sites.}
    \label{fig:core_size}
\end{figure}

\begin{figure}[H]
    \centering
    \includegraphics[width=0.9\linewidth]{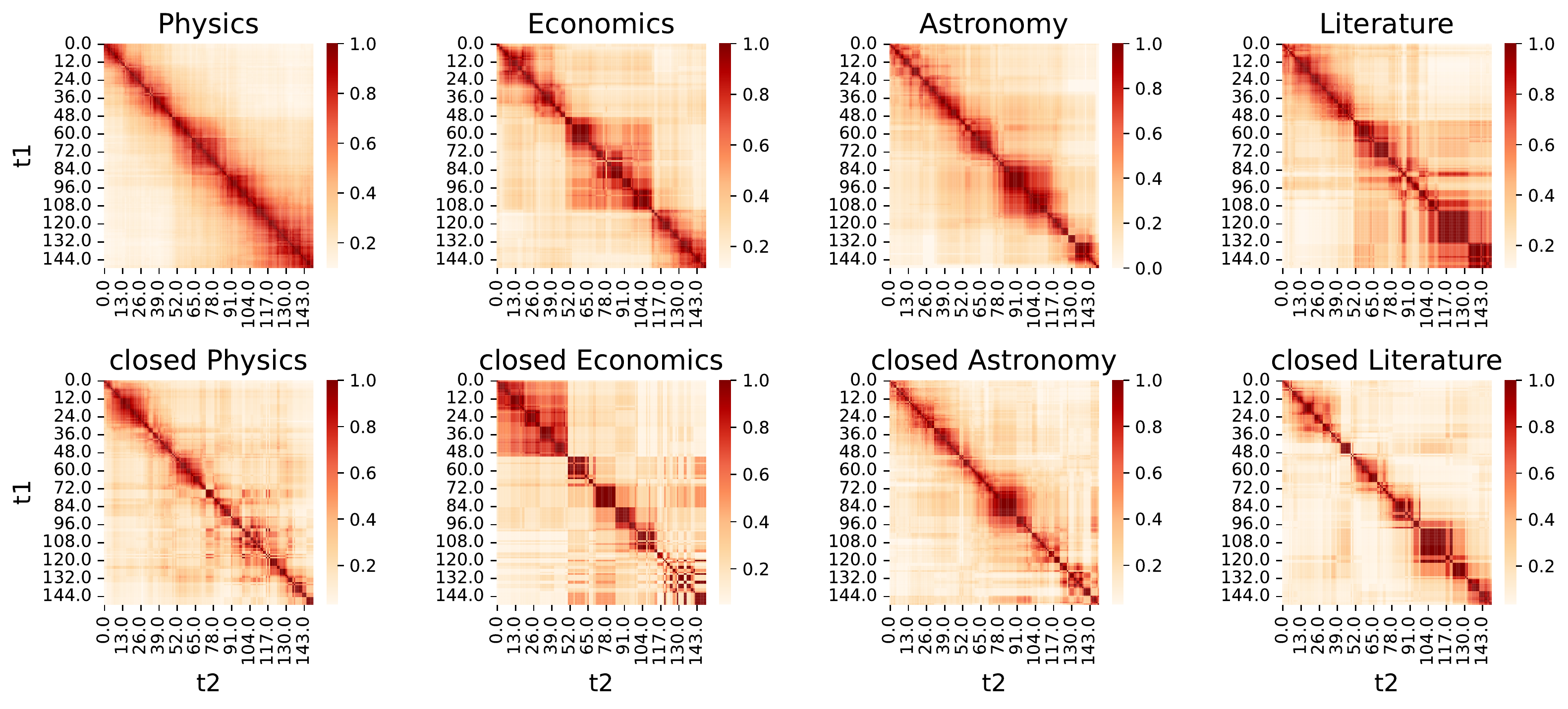}
    \caption{Jaccard index between core users in  sub-networks at time points $t1$ and $t2$}
    \label{fig:jaccard_hm}
\end{figure}

\begin{figure}[H]
    \centering
    \includegraphics[width=0.9\linewidth]{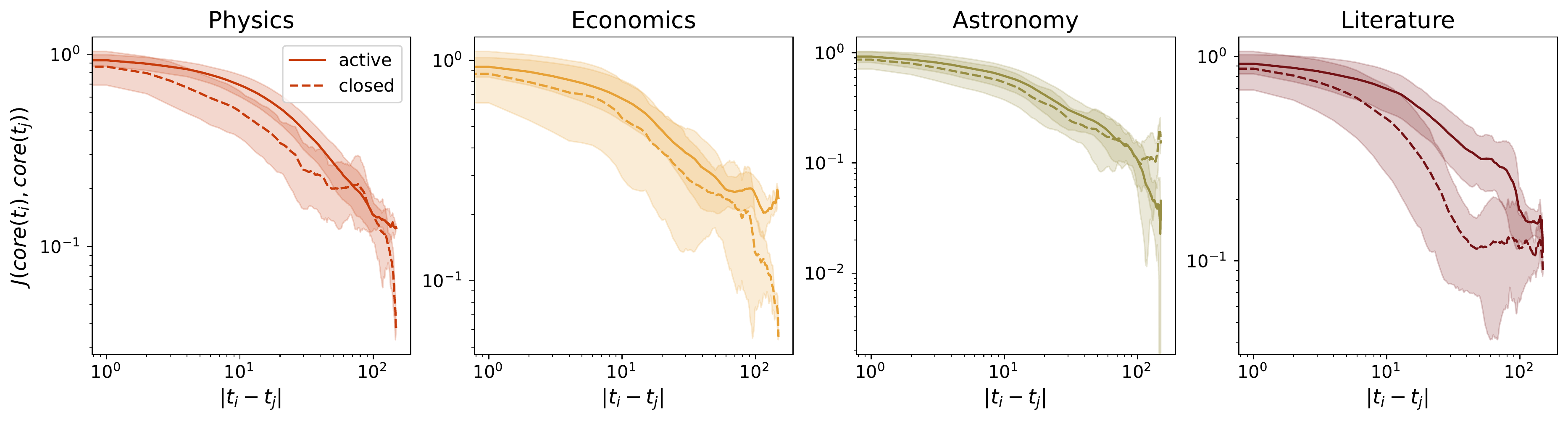}
    \caption{Jaccard index between core users in 30days sub-networks for all possible pairs of 30 days sub-networks separated by time interval $|t_i - t_j|$}
    \label{fig:jaccard_mean}
\end{figure}

To  explore the stability of the core across the time we compute Jaccard’s coefficient between core users in [t+30) networks selected at times $t_1$ and $t_2$, (figure \ref{fig:jaccard_hm}). Higher values of the Jaccard index indicate that core users tend to stay in the core. The detected cores experience a lot of change over time, and the highest overlap between core users is in the network closer in time. The average Jaccard index between core users in all subnetworks separated by time interval $|t_1 - t_2|$ with the standard deviation confidence interval is presented in Figure \ref{fig:jaccard_mean}. Compared to closed sites, active sites show more stability in the core. Even the number of core users obtained in the launched and closed communities is comparable \ref{fig:core_size} (a high difference is found only for physics ).

\newpage
\subsection{Dynamic reputation in the network of interactions}

In the few figures below, we investigate whether users' dynamic reputation is related with users' position within the network.

\begin{figure}[H]
    \centering
    \includegraphics[width=\linewidth]{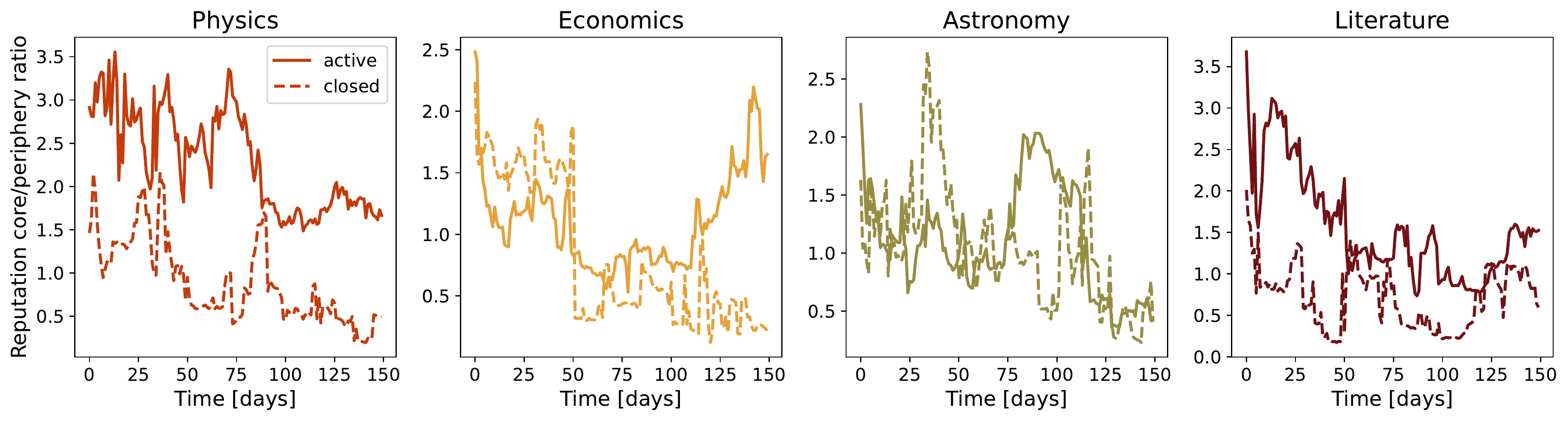}
    \caption{Ratio between the total reputation within network core and periphery. Solid lines active communities, dashed lines closed communities.}
    \label{fig:dr_core_per}
\end{figure}

The \textbf{ratio between total core and periphery reputation} is evidently higher in the active than in closed sites, figure \ref{fig:dr_core_per}.  

\begin{figure}[H]
    \centering
    \includegraphics[width=\linewidth]{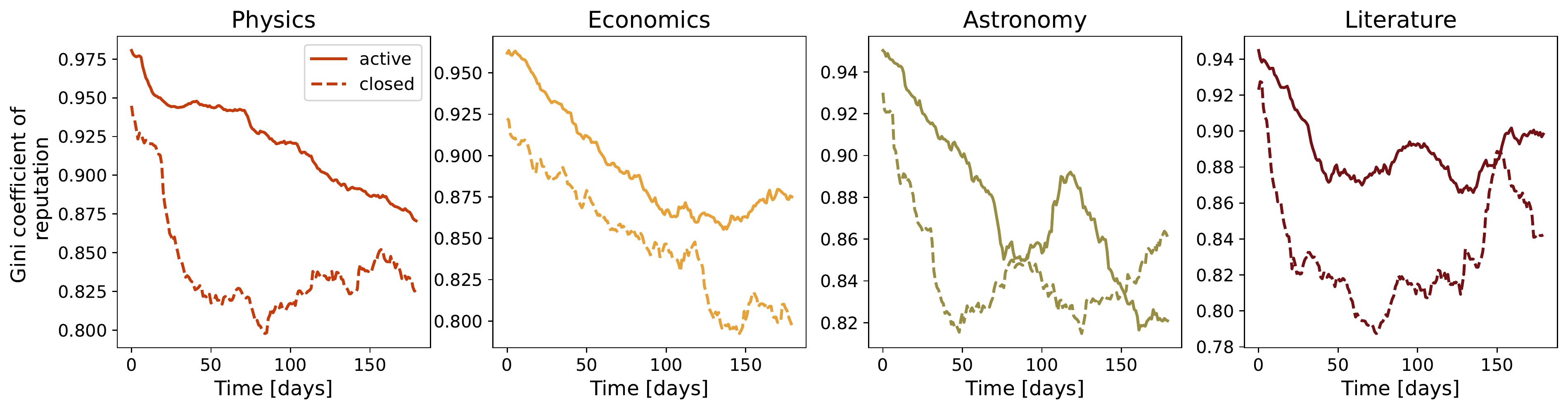}
    \caption{Gini index of dynamic reputation within population.}
    \label{fig:dynrep-gini}
\end{figure}

\textbf{Gini coefficient.} Besides the number of active users (who at a given moment of observation have a reputation higher than the threshold) and the population mean value of dynamical reputation, we have investigated in more detail the distribution of dynamical reputation within the discussed communities. We have observed that the distributions are often skewed, which prompted us to compare the communities in terms of their Gini coefficient. The Gini coefficient is a simple measure that shows us the degree of reputation inequality within the community. We calculate the value based on the dynamic reputation values of users at every time step (day) and report the values in Fig.~\ref{fig:dynrep-gini}. We see that all communities (both still active and closed ones) have Gini coefficient values higher than $0.5$ throughout the first six months. Interestingly, except in the case of Astronomy, currently active communities had higher reputation inequality every day during the first six months period. As in many other measures, in the case of astronomy, the closed community started as an unequal one (signalled by higher Gini coefficient values), but after around two months, the situation changed. 

\begin{figure}[H]
    \centering
    \includegraphics[width=\linewidth]{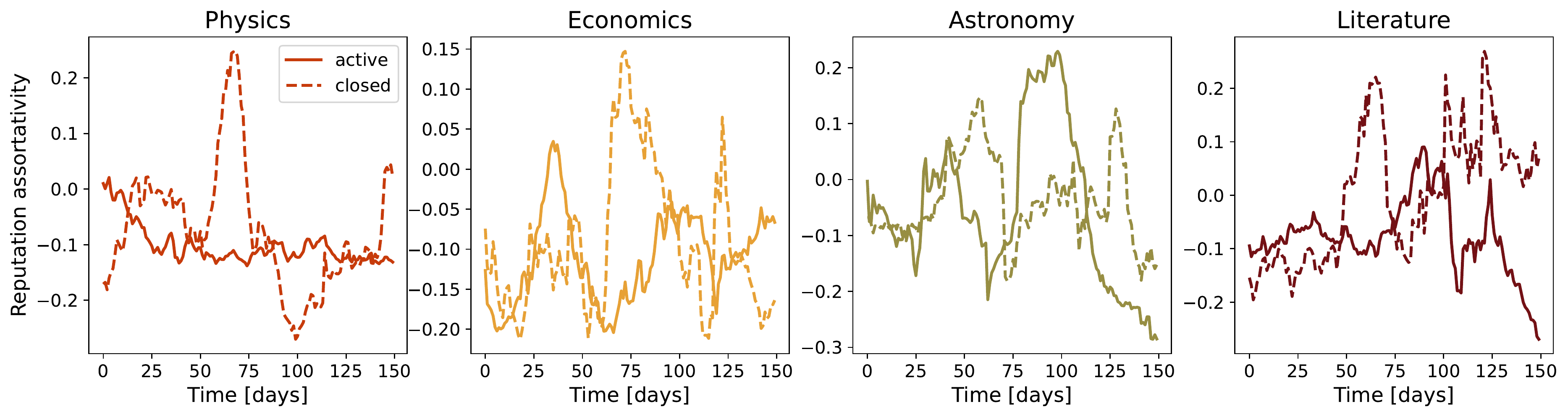}
    \caption{Dynamic Reputation assortativity within 30 days sub-networks for each community. Solid lines are active, while dashed lines are closed communities.}
    \label{fig:dyn_rep_assort}
\end{figure}

\textbf{Dynamic Reputation assortativity.} We first look at user interaction patterns; for example, we investigate whether users connect with others of similar or different reputation (positive/negative assortativity). We operationalize this by measuring the assortativity of dynamic reputation on the interaction network. Practically, this is a measure of the correlation between the dynamic reputation of users who are linked in the interaction network. These results are shown in Fig.~\ref{fig:dyn_rep_assort}. We look at 30 day unweighted undirected networks of interactions (questions, answers and comments) and calculate assortativity by using users' reputation on the last day of observed time window. We see small values of assortativity that are mostly negative, signaling weak correlations between reputation levels of interacting users. The fact that the values are mostly negative is expected, and users of different dynamic reputation interact, e.g. active, high-reputation users respond to the questions of new, less reputable users. Exceptions are closed astronomy and literature sites that occasionally had positive assortativity values, signaling existence of links between users of similar reputation levels.

\begin{figure}[H]
    \centering
    \includegraphics[width=0.9\linewidth]{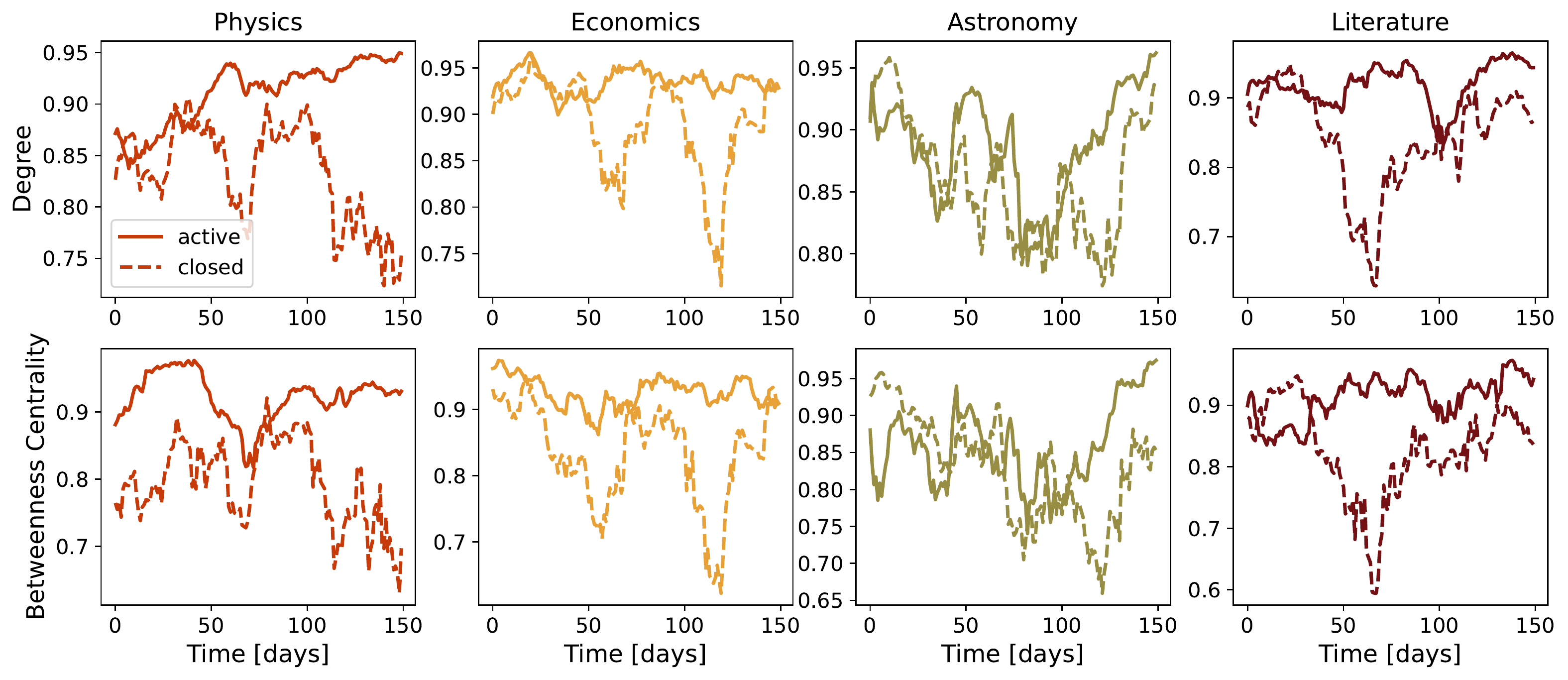}
    \caption{Coefficient of correlation between users' Dynamic Reputation and users' network degree (top) and users' betweenness centrality (bottom). Solid lines - active sites; dashed lines - closed sites.}
    \label{fig:dyn_rep_centrality}
\end{figure}

\textbf{Dynamic Reputation \& network centrality measures.} We continue to investigate whether the user's reputation correlates with typical network centrality measures calculated at user's node in the interaction network. Particularly, we investigate degree and betweenness centrality measures.
As previously, we compare the node's centrality in the 30 day network with the node's dynamic reputation on the last day of the period, repeating the process every day for the first six months. 
The correlation coefficient between dynamic reputation and degree in the network is very high, as expected, as most of the interactions that contributed to user's reputation are also present as links in the network. We show these results in Fig.~\ref{fig:dyn_rep_centrality} (top). However, we again see the distinction between active and closed communities, where this correlation is higher in active communities, except in the first month of sliding windows. Astronomy is an exception here as well as we see that the correlations were similar in both closed and still active sites throughout the observed period. 
In the bottom panels of Fig.~\ref{fig:dyn_rep_centrality} we present correlation coefficients of dynamic reputation and user's betweenness centrality in the interaction network. These correlations are also high and, most of the time, higher in the later networks of active than closed communities. This is particularly interesting due to the global nature of the betweenness centrality measure and less obvious relation to the user's dynamic reputation.

\subsection{The choice of the sliding window}

\begin{figure}[H]
    \centering
    \includegraphics[width=0.78\linewidth]{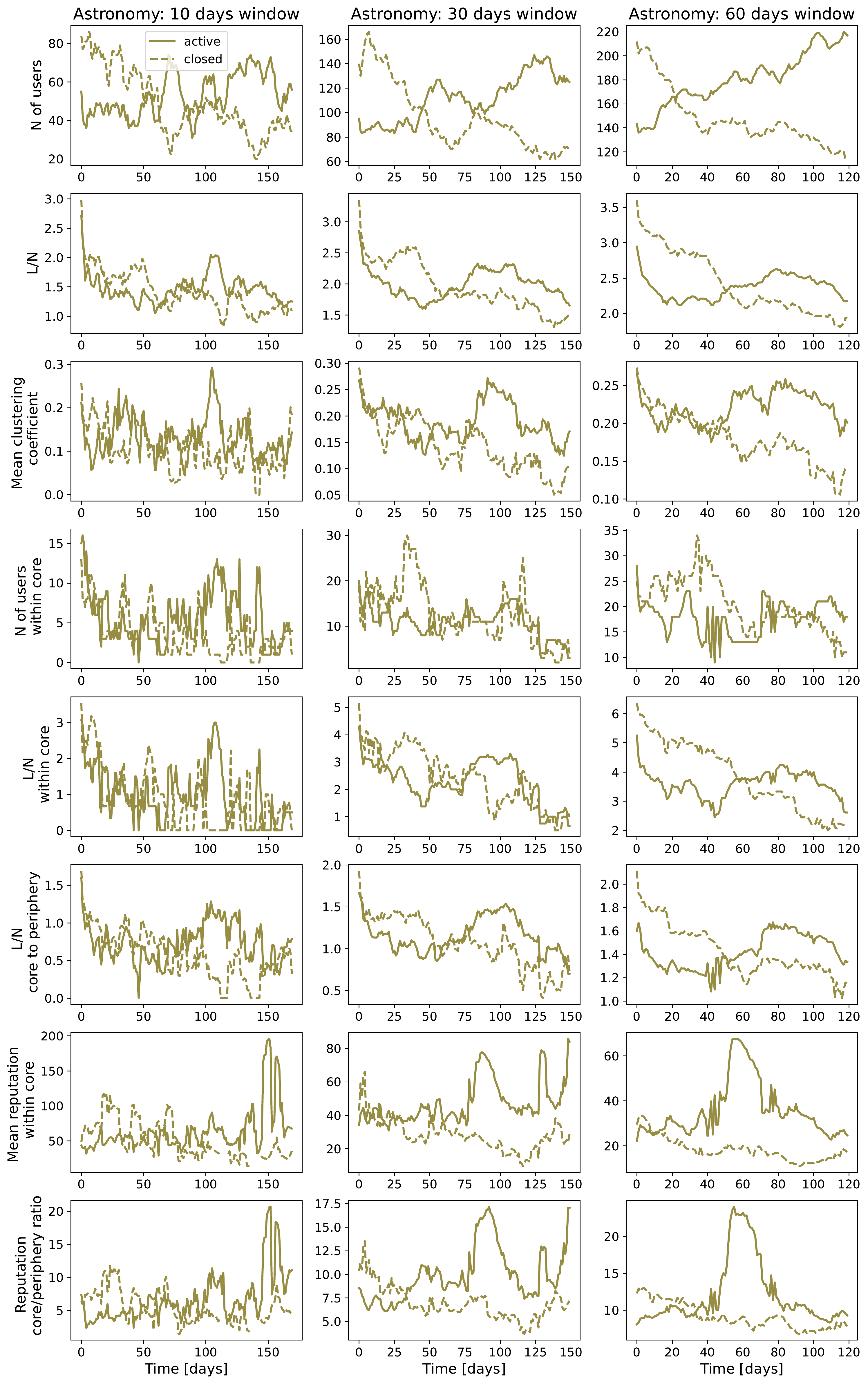}
    \caption{The comparison of network and core-periphery measures within sub-networks for different choice of sliding window $\tau = [10, 30, 60]$. Solid lines - Active Astronomy, dashed lines - Closed Astronomy. }
    \label{fig:windows}
\end{figure}

In this section, we investigate how the size of sliding windows affects network properties over time. Figure \ref{fig:windows} summarizes the results for one pair of communities, closed and active astronomy, but similar conclusions can be observed for other pairs of sites. We show the network properties for subnetworks of 10, 30, and 60 days sliding windows. For a sliding window of 10 days, results may be too noisy, and we may not observe some important trends in the community. The number of users for active astronomy seems to fluctuate around some mean value. On a larger scale, with the 30-day window, it is more clear that the number of users increases slightly over time. On the contrary, for too large an aggregation window (60 days), important information about the time series can be lost, such as the local minimum of the number of users around time step 80 that is observed for the 30-day sliding window. Looking at other network characteristics, such as L / N and clustering, we conclude that the differences between closed and active sites are more transparent with a larger aggregation window; still, on each scale, beta sites show a higher number of nodes, the number of links per node, and the clustering coefficient.
As before, we study the structure of created subnetworks through the lens of the core-periphery structure. On small scales, within the window of 10 days, there are often few or even no nodes in the core, and it can affect the calculation of other measures of interest. This behavior is more typical for closed communities.  With the size of the sliding window number of nodes in the core increases, and the results of core-periphery measures become smoother. Finally, the choice of the sliding window does not change the conclusion that core users in the beta communities produce more activity and make a strong core. However, our main results are shown for a sliding window of 30 days, as it makes a good compromise between large and small time scales. \\ 

\clearpage
\newpage
\subsection{Model parameters}

\begin{figure}[H]
    \centering
    \includegraphics[width=\linewidth]{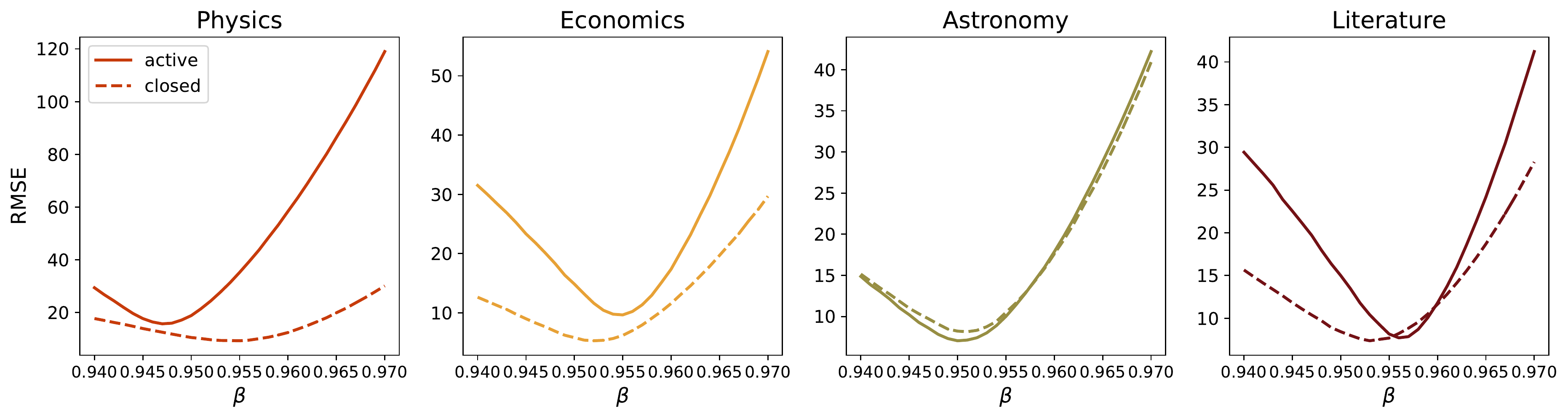}
    \caption{Root mean squared error (RMSE) between number of users in sliding window of 30 days and number of users with reputation higher than $1$ for $0.94< \beta <0.97$ with step $0.001$. }
    \label{fig:rmse}
\end{figure}

We compared the number of users with an estimated reputation greater than 1 for different parameters $\beta$ and concluded that $\beta$ close to $0.96$ approximates the number of users with recorded interactions in a given 30-day sliding window. For each pair of communities, we calculated the number of users with at least one interaction in every 30 day sliding window. Then, we estimated several times series expressing the number of users with reputation higher than 1 for fixed $\beta$. Then we calculated the root mean square error (RMSE) between those time series for the first 180 days. RMSE values are shown in Figure~\ref{fig:rmse}. For each community, we can find the parameter $\beta$ that minimizes RMSE. Although $\beta$ does not have a unique value across communities, it varies between 0.95 and 0.96.

\begin{figure}[H]
    \centering
    \includegraphics[width=\linewidth]{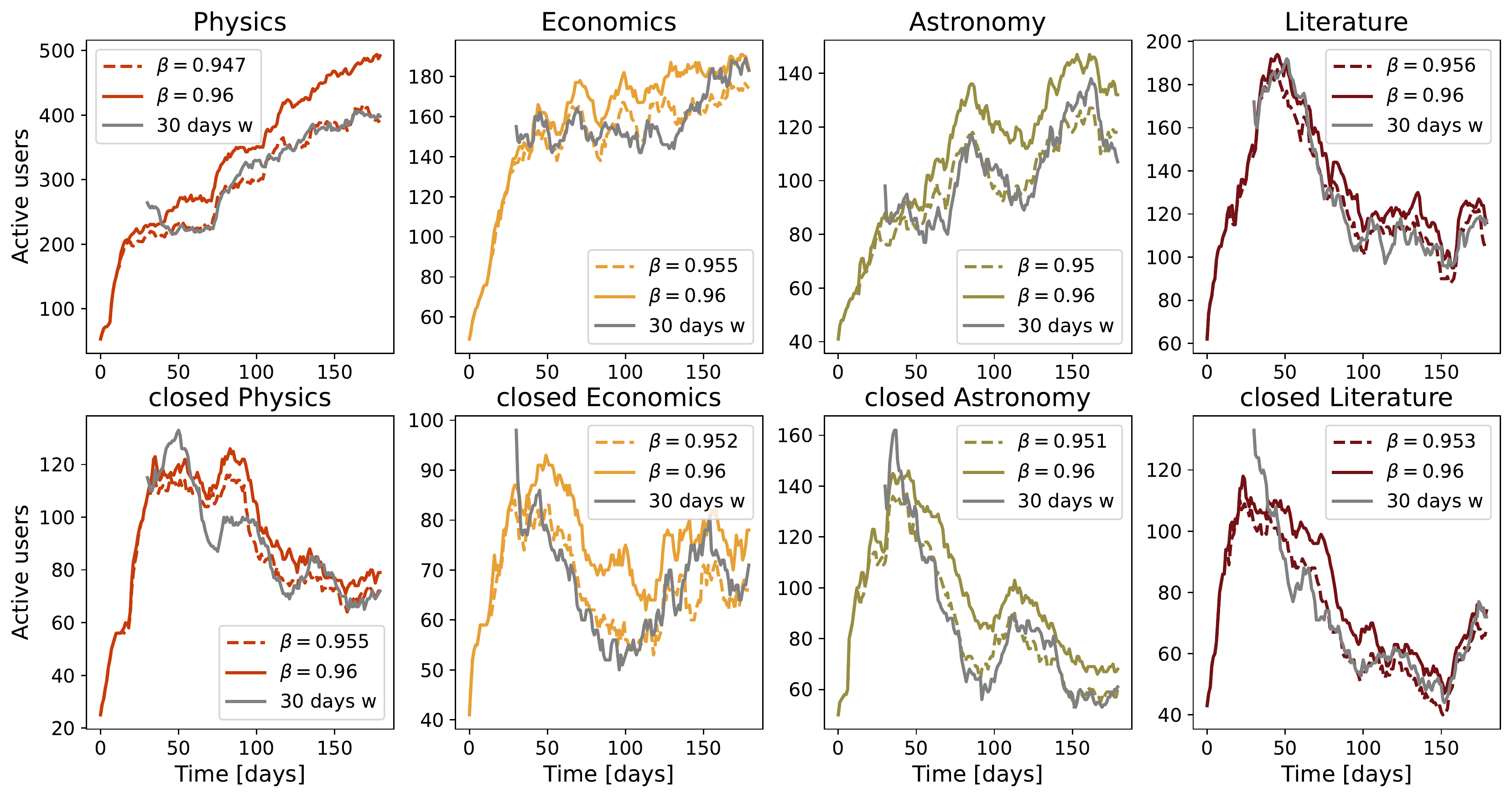}
    \caption{Number of users with dynamic reputation higher than 1 for $\beta=0.96$ and $\beta$ which provide the best fit to the number of users in 30 days sub-networks for each community. For the reference we plot the number of active users within 30 days sliding window.}
    \label{fig:nusers}
\end{figure}

Figure \ref{fig:nusers} shows comparison between number of users in 30 days sliding window, number of users for these optimal values $\beta$ and $\beta =0.96$. For $\beta = 0.96$, we observe that in most communities, the estimated number of active users is consistently slightly higher than the actual number of users who have made at least one interaction in that sliding window. This means that the dynamic reputation model, in some cases, overestimates user's reputation,, but much more important is that it never underestimates the real number of active users. Since we base our calculations of total and average reputation within the community only on users whose reputation is higher than the threshold, this is important because the model disregards no active users due to the value of the decay parameter.

\begin{figure}[H]
    \centering
    \includegraphics[width=1\linewidth]{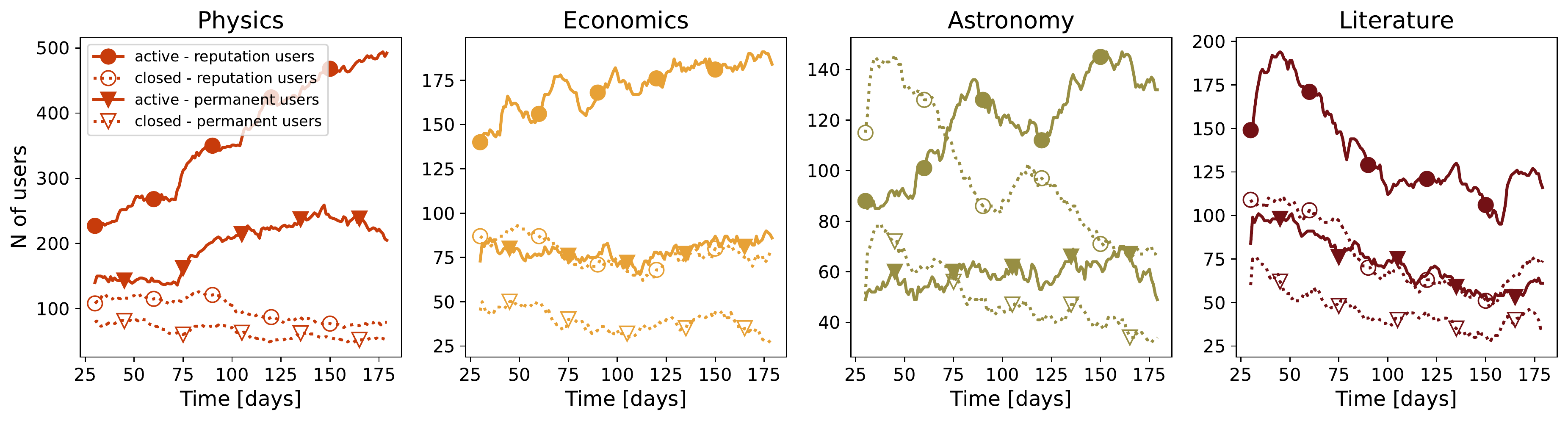}
    \caption{The number of users with dynamic reputation higher than 1 for $\beta=0.96$ (reputation users - circles) and number of users within 30 days sliding window who had activity in current time window and remained to be active in the following time windows (permanent users - triangles). Solid lines are active while dashed lines are closed communities.  }
    \label{fig:active-users}
\end{figure}

Finally, it is important that our dynamic reputation captures the trend of long-term user activity. In Figure~\ref{fig:active-users} lines with circles show the time series of estimated dynamic reputation for $\beta = 0.96$ while lines with triangles show the number of users who were active in a given sliding window and continued to be active in the next one. Although the total estimated number of active users is expected to be higher, two time series follow similar trends in different communities.

\end{document}